\documentclass[12pt]{article}

\usepackage{color,extarrows} 
\numberwithin{equation}{section}

\newcommand{\be}{\begin{equation}}
\newcommand{\ee}{\end{equation}}
\usepackage{amsmath}
\usepackage{amssymb}
\usepackage{latexsym}
\usepackage{epsfig}
\usepackage{multirow}
\usepackage{lscape}

\newcommand{\bea}{\begin{eqnarray}}
\newcommand{\eea}{\end{eqnarray}}

\newcommand{\RRR}{{\hbox{\rm R\kern-2.35mm R}}}

\def\ZZZ{{\hbox{ Z\kern-1.6mm Z}}}

\begin{document}

\begin{titlepage}
\rightline{August 2019}
\rightline{  Imperial-TP-2019-CH-06}
\begin{center}
\vskip 2.5cm
{\Large \bf {
Special Holonomy Manifolds, Domain Walls, Intersecting Branes and T-folds}}\\
\vskip 2.0cm
{\large {N.  Chaemjumrus and C.M. Hull  }}
\vskip 0.5cm
{\it {The Blackett Laboratory}}\\
{\it {Imperial College London}}\\
{\it {Prince Consort Road}}\\
{\it { London SW7 @AZ, U.K.}}\\

\vskip 2.5cm
{\bf Abstract}
\end{center}

\vskip 0.5cm

\noindent
\begin{narrower}
We discuss the special holonomy metrics of
Gibbons, Lu, Pope and Stelle, which were constructed  as nilmanifold bundles over a line by uplifting supersymmetric domain wall solutions of supergravity to 11 dimensions.
We show that these are dual to intersecting brane solutions, and considering these leads us to a more general class of     special holonomy metrics.
Further dualities relate these to non-geometric backgrounds involving intersections of branes and exotic branes.
We discuss the possibility of resolving  these spaces to give smooth special holonomy manifolds.

\end{narrower}

\end{titlepage}

\newpage

\tableofcontents
\baselineskip=16pt
\section{Introduction}

The nilfold is a compact 3-manifold that is the quotient of the group manifold of the Heisenberg group by a discrete subgroup. It is a circle bundle over a 2-torus and its T-duals include a $T^3$ with $H$-flux~\cite{Hull:1998vy}
  and a T-fold~\cite{Kachru:2002sk,Hull:2004in}.
It can be incorporated into string theory by fibring over a line: a nilfold bundle over a line admits a hyperk\" ahler metric and so gives a half-supersymmetric string background~\cite{Hull:1998vy, Lavrinenko:1997qa,Gibbons:1998ie}.
The metric is of Gibbons-Hawking type \cite{Gibbons:1979xm}, given by a harmonic function which is a piecewise linear function on the line, with the kinks at which there are jumps in the gradient associated with  domain walls.
T-dualising gives a  $T^3$ with $H$-flux fibred over a line \cite{Hull:1998vy}, which can be viewed as a multi-NS5-brane solution wrapped on $T^3$. This in turn is dual~\cite{Hull:1998vy} to a multi-D8-brane solution \cite{Bergshoeff:1996ui}.
Again this is governed by a piecewise linear function on the line, with the kinks associated with domain walls that are  D8-branes wrapped on $T^3$ or NS5-branes  smeared over a transverse $T^3$.
It is also T-dual to a T-fold fibred over a line~\cite{Hull:2004in,Ellwood2006,Chaemjumrus:2019ipx}. In
 \cite{Chaemjumrus:2019ipx}, the incorporation of these solutions into complete consistent string backgrounds was discussed.

Gibbons, Lu, Pope and Stelle showed that this hyperk\" ahler space has a remarkable generalisation to higher dimensions in which the hyperk\" ahler metric on the nilfold fibred over a line is generalised to a special holonomy metric on a higher dimensional nilmanifold $M$
fibred over a line \cite{Gibbons:2001ds}.
Replacing the Heisenberg group by a higher dimensional nilpotent lie group and quotienting by a suitable discrete subgroup gives a compact manifold which we will refer to as a nilmanifold.
In the simplest cases that we will focus on here, this is a $T^n$ bundle over $T^m$ for
some $m,n$.  More generally it can be   a finite series of torus bundles: starting with a torus $T^{n_1}$, a 
$T^{n_2}$ bundle is constructed over this, then a $T^{n_3}$ bundle is constructed over this bundle, then a $T^{n_4}$ bundle is constructed over this, and so on for some finite set of integers $n_1,n_2,\dots , n_r$.
It was shown in \cite{Gibbons:2001ds} that 
fibring certain nilmanifolds over a line gives manifolds of special holonomy. This result  was found by considering supersymmetric domain walls in $D$-dimensional supergravity that could be uplifted to compactifications of 11-dimensional supergravity on a nilmanifold $M$ of dimension $n=11-D$, with trivial 3-form gauge field. Such solutions are a product of $D-1$ dimensional Minkowski space with a space which is $M$ fibred over the line transverse to the domain walls.  As the $D$-dimensional solution is supersymmetric, the 11-dimensional one is also supersymmetric, which implies that the $M$ bundle over the line has special holonomy. Spaces of holonomy $SU(3),SU(4),G_2,Spin(7)$ were constructed in this way.
The results of \cite{Gibbons:2001ds}   have been extended  to a larger class of nilmanifolds, and   special holonomy metrics
on these nilmanifolds fibred over a line have been found in \cite{Chiossi and Salamon(2002),Conti(2009), Apostolov(2004), Bruun(2018)}.

Here we will extend some of the analysis of \cite{Chaemjumrus:2019ipx} for the nilfold case to the solutions of \cite{Gibbons:2001ds};  we expect similar results will apply
to  the special holonomy spaces of 
\cite{Chiossi and Salamon(2002),Conti(2009), Apostolov(2004), Bruun(2018)}. The hyperk\" ahler solution from the nilfold is dual to wrapped D8-branes or smeared NS5-branes, 
We will show here that each of the special holonomy spaces is dual to a system of intersecting branes wrapped on a torus that preserves precisely the same amount of supersymmetry.
This in turn leads us to a modest generalisation of the special holonomy metrics of \cite{Gibbons:2001ds}, which involved one piecewise linear function: the general intersecting brane solution allows different harmonic functions for each brane, and dualising this leads to special holonomy metrics specified by several piecewise linear functions.
We will then consider  T-dualities to non-geometric solutions involving T-folds fibred over a line. Further T-dualities give essentially doubled non-geometric solutions (with R-flux) that can only be presented in a doubled geometry;  these will be discussed in a seperate paper \cite{ChaemHull}.

These multi-domain wall solutions are typically singular at the locations of the  domain walls.
For the nilfold case, the singular four dimensional solution has a remarkable resolution to give a complete smooth hyperk\" ahler metric, provided the total brane charge is not greater than 18 \cite{2018arXiv180709367H}.
The result is a K3 metric, in a region of moduli space in which the K3 has a long neck for which the smooth metric is well approximated by the 
multi-domain wall metric away from the domain wall singularities. 
K3 metrics of precisely this type were constructed in \cite{2018arXiv180709367H}.
The neck is essentially a nilfold fibred over a line, with  Kaluza-Klein monopole \lq bubbles' inserted at a series of points on the neck that provide the resolution of the domain wall singularities. At either end of the neck is a cap, which is obtained by glueing on a Tian-Yau space~\cite{Tian1990}, which is a non-compact hyperk\" ahler space that is asymptotoic to a nilfold fibred over a line.

The existence of  K3 metrics of this form could have been anticipated by an argument based on string dualities, as discussed in \cite{Chaemjumrus:2019ipx}.
Type I$'$ string theory has a consistent solution which is a product of 9-dimensional Minkowski space with a finite interval, with an O8 orientifold plane at either end of the interval and 16 D8 branes at arbitrary points along the interval. Strictly speaking, this is the picture at weak coupling; at strong coupling there can be up to 18 D8 branes \cite{Morrison:1996xf,Gorbatov:2001pw}.
Dualising this should give a consistent solution of the corresponding theory dual   to type I$'$ string theory.
Compactifying on $T^3$ and T-dualising takes the D8 branes to D5 branes and the O8 planes to O5 planes.
S-duality   takes this to NS5 branes and ON-planes. Then a further T-duality takes the NS5-branes to Kaluza-Klein monopoles. 
This same chain of dualities takes the type I$'$ string theory on $T^3$ to the type IIA string theory on $T^4/\mathbb{Z} _2$, which is an orbifold limit of type IIA string theory on K3. Moving in the type I$'$ string theory moduli space should then translate to moving in the moduli space of  type IIA string theory on K3. A long interval in type I$'$  translates into a K3 with a long neck, and moving D8 branes on the interval translates into moving Kaluza Klein bubbles on the neck. Strong coupling in the type I$'$ theory is mapped to weak coupling in the type IIA dual, and 
so there can be up to 18 Kaluza Klein monopoles.
The O8 planes are mapped to the Tian-Yau spaces, and it is the classification of these spaces, which are obtained  from   del Pezzo surfaces,  that restricts the total charge to be no greater than 18.

We expect similar arguments to apply to the special holonomy spaces obtained here. The brane intersections that arise as their duals can be incorporated into consistent string backgrounds, and the dualities should take the backgrounds to new consistent string backgrounds, which should be smooth compact special holonomy manifolds.
Thus it is to be expected that there should be generalisations of the K3 construction of \cite{Chaemjumrus:2019ipx,2018arXiv180709367H} to special holonomy spaces.
The K3 metric studied in \cite{2018arXiv180709367H} arises near a boundary of moduli space in which the K3 is  degenerating
to a line segment. The corresponding special holonomy solutions would arise in similar degenerating limits.

In \cite{Sun}, possible extensions of the K3 construction of  \cite{2018arXiv180709367H} to higher dimensional Calabi-Yau spaces were considered, using neck regions in which the Gibbons Hawking ansatz of  \cite{2018arXiv180709367H} was replaced by the non-linear ansatz introduced in \cite{Zha} and recently developed in \cite{Li}. 
The 4-dimensional   hyperk\" ahler Tian-Yau  spaces, constructed by removing an elliptic curve from a del Pezzo surface, 
 used as the caps in \cite{2018arXiv180709367H}, were replaced in 
\cite{Sun} by higher dimensional Tian-Yau  spaces that are asymptotically cylindrical non-compact  Calabi-Yau spaces  and constructed from Fano spaces instead of del Pezzo surfaces.
However, the non-linearity of the construction of  \cite{Zha} made this complicated to use.
We propose  to instead use the Calabi-Yau and special holonomy metrics discussed here as model geometries for the neck region. This has the advantage that the ansatz is linear, making the analysis easier. The domain wall singularities should be resolvable in a similar way to those of \cite{2018arXiv180709367H}.
The local geometry would  then be a nilmanifold fibred over a line, and for the cap regions a Tian-Yau type space which is Calabi-Yau or special holonomy and asymptotic for 
a nilmanifold fibred over a line is  needed.


\section{The Nilfold and its T-duals}

The nilfold ${\cal N}$ is an $S^1$ bundle over a 2-torus where the 2-torus has   coordinates $x,z$ while the fibre coordinate is $y$, and has metric
\begin{equation}
ds^2_{\cal N}=dx^2+(dy-mxdz)^2+dz^2   \qquad  H=0 \label{nilfold}
\end{equation}
where the integer $m$ is the Chern number and  is referred to as the degree of the nilfold.
 The global structure of the nilfold is given by  the following identifications of the local
coordinates 
\begin{equation}\label{delta}
(x,y,z)\sim (x+1,y+mz,z)    \qquad  (x,y,z)\sim (x,y+1,z)   \qquad  (x,y,z)\sim (x,y,z+1).
\end{equation}
Locally, it is a group manifold.  The 3-dimensional Heisenberg group $G_3$ has 
Lie algebra
$$
[T_x,T_z]=mT_y  \qquad  [T_y,T_z]=0 \qquad [T_x,T_y]=0
$$
generated by upper triangular $3\times 3 $ matrices.
 Using  local coordinates $(x,y,z)$ on the group manifold $G_3$, 
 a general  group element can be written using    coordinates $(x,y,z)$ as
\begin{eqnarray}\label{g}
g=\left(%
\begin{array}{ccc}
  1 & mx & y \\
  0 & 1 & z \\
  0 & 0 & 1 \\
\end{array}%
\right).
\end{eqnarray}
  It has a  discrete subgroup $\Gamma$  of matrices
  \begin{eqnarray}\label{h}
h=\left(%
\begin{array}{ccc}
  1 & m\alpha & \beta \\
  0 & 1 & \gamma \\
  0 & 0 & 1 \\
\end{array}%
\right)
\end{eqnarray}
with $\alpha$, $\beta$ and $\gamma$   integers. 
 Then the nilfold is given by the quotient of $G_3$ by the discrete subgroup $\Gamma$.
 The nilfold can also be viewed as a 2-torus bundle over a circle, with a 2-torus parameterised by $y,z$ and base circle parameterised by $x$.

T-dualising in the $y$ direction gives a 3-torus with $H$-flux given by an integer $m$. The metric and 3-form flux $H$ are
\begin{equation}\label{gH}
ds^2_{T^3}=dx^2+dy^2+dz^2   \qquad  H=mdx\wedge dy\wedge dz
\end{equation}
with  periodic coordinates
$$
x\sim x+1  \qquad  y\sim y+1 \qquad  z \sim  z+1.
$$
Here flux quantisation requires that  $m$ is an integer.

T-dualising the nilfold in the $z$ direction gives
the T-fold
with
metric and $B$-field given by
\begin{equation}\label{gb}
ds^2_{T-fold}=dx^2+\frac{1}{1+(mx)^2}(dy^2+dz^2)   \qquad  B=-\frac{mx}{1+(mx)^2}dy\wedge dz 
\end{equation}
which changes by a T-duality under $x\to x+1$, and so
has a T-duality monodromy in the $x$ direction.
A further T-duality in the $x$ direction gives something which is not locally geometric but which has a well-defined doubled geometry given in \cite{Hull:2009sg,ReidEdwards:2009nu}.

\section{Nilmanifolds as Torus Bundles over Tori} \label{Nilfolds}

\subsection{Nilpotent Groups and Nilmanifolds }

In this subsection, we review certain generalizations of the 3-dimensional  nilfold    to higher dimensions. The Heisenberg group is replaced by a higher dimensional nilpotent Lie group $\mathcal{G}$, and taking the quotient by a cocompact discrete subgroup gives a 
nilmanifold, which is a 
compact space which is a $T^n$ bundle over $T^m$ for some $m,n$.

  A Lie algebra $\mathfrak{g}$   is  nilpotent  if the lower central series terminates, that is
\begin{equation}
[X_1,[X_2,[\cdots[X_p,Y]\cdots]] = 0
\end{equation} 
for all $X_1, \cdots, X_p, Y \in \mathfrak{g}$, for some integer  $p $. For a nilpotent Lie group $\mathcal{G}$, the smallest  such $p$ is known as the nilpotency class of $\mathcal{G}$ and $\mathcal{G}$ is then called a $p$-step nilpotent Lie group. Note that an abelian group is a 1-step nilpotent Lie group since 
\begin{equation}
[X,Y] = 0
\end{equation}
for all $X, Y \in \mathfrak{g}$. The 3-dimensional Heisenberg group is a 2-step nilpotent Lie group since
\begin{equation}
[T_x,T_z] = mT_y
\end{equation}
and $T_y$ commutes with $T_x$ and $T_z$.

For a general 2-step nilpotent Lie group $\mathcal{G}$, the commutator of any two generators $X,Y$ of   the Lie algebra $\mathfrak{g}$
must be in the centre  $Z(\mathfrak{g})$ of $\mathfrak{g}$ (consisting of generators commuting with all other generators):
\begin{equation}
[X,Y] \in Z(\mathfrak{g}) \, .
\end{equation}
In general, a 2-step nilpotent Lie group $\mathcal{G}$ is non-compact. A compact space $\mathcal{M}$ can be constructed from a nilpotent Lie group $\mathcal{G}$ identified under the left action of a cocompact subgroup $\Gamma$:
\begin{equation}
\mathcal{M} = \mathcal{G}/\Gamma.
\end{equation}
If the dimension of $Z(\mathfrak{g})$  is $n$ and the dimension of the quotient $\mathfrak{g}/Z(\mathfrak{g})$ is $m$, the compact space $\mathcal{M}$ can be regarded   as  a $T^n$ bundle over $T^m$. For example, the Heisenberg group $G_3$ has 
a centre $Z(\mathfrak{g})$ of dimension $n=1$ and the  dimension of $\mathfrak{g}/Z(\mathfrak{g})$  is $m=2$, and the
nilfold is indeed an $S^1$ bundle over $T^2$.

Local coordinates $z^a$ on the group manifold $\mathcal{G}$ can be introduced by the exponential map giving a group element $g$ as
$g =\displaystyle\prod_a\exp (z^aT_a)$ where
 $T_a$ are the Lie algebra generators.
The  general left-invariant metric on $\mathcal{G}$ can be written as
\begin{equation}
ds^2 = x_{ab}P^aP^b,
\end{equation}
where $x_{ab}$ is a constant symmetric matrix and  $P^a$ are the left-invariant one-forms
\begin{equation}
g^{-1}dg = P^aT_a .
\end{equation} 
In this paper, $x_{ab}$ will be chosen as $x_{ab} = \delta_{ab}$ so the left-invariant metric is
\begin{equation}
\label{metr}
ds^2 = \delta_{ab}P^aP^b.
\end{equation}
The discrete subgroup $\Gamma$
consists of group elements with integer coordinates,  $g(n) = \displaystyle\prod_a\exp (n^aT_a)$ for integers $n^a$.
Taking the quotient of $\mathcal{G}$ by the left action of the discrete subroup $\Gamma$ gives the nilfold $\mathcal{M} = \mathcal{G}/\Gamma $ and (\ref{metr}) gives a metric on $\mathcal{M} $.
Taking the quotient   imposes identifications on the coordinates so that the space becomes a torus bundle over a torus.

We now consider the explicit examples that arise in \cite{Gibbons:2001ds}.

\subsection{$S^1$ bundle over $ T^4$}

Our first example is the five-dimensional nilpotent Lie algebra whose  only non-vanishing commutators are
\begin{equation}
[T_2,T_3] = mT_1,\qquad [T_4,T_5] = mT_1.
\end{equation}
With coordinates $z^1,\dots, z^5$, 
the left-invariant one-forms are   given by
\begin{eqnarray}
P^1 &=& dz^1+m(z^3dz^2+z^5dz^4),\nonumber\\
P^2 &=& dz^2,\;
P^3 \;=\; dz^3,\nonumber\\
P^4 &=& dz^4,\;
P^5 \;=\; dz^5.
\end{eqnarray}
The metric (\ref{metr}) is then
\begin{equation}
ds^2 = \Big(dz^1+m(z^3dz^2+z^5dz^4)\Big)^2+(dz^2)^2+(dz^3)^2+(dz^4)^2+(dz^5)^2. \label{T1T4metric}
\end{equation}
This is an $S^1$ bundle over $T^4$ with fibre coordinate $z^1$ and $T^4$ coordinates $z^2,z^3,z^4,z^5$, with first Chern class represented by
$$ F= m\Big(dz^3 \wedge dz^2+dz^5 \wedge dz^4\Big){.}
$$
The metric is invariant under shifts of  $z^1$, $z^2$, and $z^4$ so that T-dualising in these directions is straightforward, applying the standard Buscher rules. We will not give all dual backgrounds explicitly, but focus on some interesting examples.

T-duality in the $z^1$ direction gives a $T^5$ with $H$-flux. The metric and $H$-flux of this space are given by
\begin{eqnarray}\label{t5h}
ds^2 &=& (dz^1)^2+(dz^2)^2+(dz^3)^2+(dz^4)^2+(dz^5)^2,\\
H &=& -mdz^1\wedge dz^2 \wedge dz^3 - mdz^1 \wedge dz^4 \wedge dz^5 {.} \label{t5hh}
\end{eqnarray}

T-dualising the metric (\ref{T1T4metric})  in the  $z^2$ and $z^4$ directions gives a T-fold background. The metric and $B$-field are given by
\begin{eqnarray}
ds^2 &=& \frac{1}{1+m^2\Big[(z^3)^2+(z^5)^2\Big]}\Big((dz^1)^2+(dz^2)^2+(dz^4)^2\Big)\nonumber\\
& &+ \frac{1}{1+m^2\Big[(z^3)^2+(z^5)^2\Big]}\Big(mz^5dz^2-mz^3dz^4\Big)^2+(dz^3)^2+(dz^5)^2 {,}\\
B &=& \frac{m}{1+m^2\Big[(z^3)^2+(z^5)^2\Big]}\Big(z^3dz^1\wedge dz^2+z^5 dz^1\wedge dz^4\Big) {.}
\end{eqnarray}
This has T-duality monodromies in the $z^3$ and $z^5$ directions.

\subsection{$T^2$ bundle over $T^3$}

Next consider the five-dimensional nilpotent Lie algebra whose  only non-vanishing commutators are
\begin{equation}
[T_3,T_4] = mT_1,\qquad [T_3,T_5] = mT_2.
\end{equation}
Introducing coordinates $z^1,\dots, z^5$,
the left-invariant one-forms are given by
\begin{eqnarray}
P^1 &=& dz^1+mz^4dz^3,\nonumber\\
P^2 &=& dz^2+mz^5dz^3,\nonumber\\
P^3 &=& dz^3,\;
P^4 \;=\; dz^4,\;
P^5 \;=\; dz^5.
\end{eqnarray}
and the metric on the manifold is
\begin{equation}
ds^2 = \Big(dz^1+mz^4dz^3\Big)^2+\Big(dz^2+mz^5dz^3\Big)^2+(dz^3)^2+(dz^4)^2+(dz^5)^2. \label{T2T3metric}
\end{equation}
This space  is a $T^2$ bundle over $T^3$
with fibre coordinates $z^1, z^2$.

T-duality in the  $z^1$  direction gives an $S^1$ bundle over $T^4$ with $H$-flux;
the metric and $H$-flux are
\begin{eqnarray}
ds^2 &=& (dz^1)^2+\Big(dz^2+mz^5dz^3\Big)^2+(dz^3)^2+(dz^4)^2+(dz^5)^2,\\
H &=& - mdz^1 \wedge dz^3 \wedge dz^4 {.}
\end{eqnarray}
A further T-duality  in the $z^2$  direction gives $T^5$ with $H$-flux. The metric and $H$-flux of this space are
\begin{eqnarray}
ds^2 &=& (dz^1)^2+(dz^2)^2+(dz^3)^3+(dz^4)^2+(dz^5)^2,\\
H &=& -mdz^1\wedge dz^3\wedge dz^4-mdz^2\wedge dz^3 \wedge dz^5.
\end{eqnarray}
After a change of coordinates, this is the same solution as 
(\ref{t5h}), (\ref{t5hh}), establishing that the $S^1$ bundle over $T^4$ is T-dual to the $T^2$ bundle over $T^3$.

Starting from the metric (\ref{T2T3metric}) and doing  a T-duality in $z^3$  direction gives  a T-fold with metric and $B$-field
\begin{eqnarray}
ds^2 &=& \frac{1}{1+m^2\Big[(z^4)^2+(z^5)^2\Big]}\Big[(dz^1)^2+(dz^2)^2+(dz^3)^2\Big]\nonumber\\
&&+\frac{1}{1+m^2\Big[(z^4)^2+(z^5)^2\Big]}\Big(z^5dz^1-z^4dz^2\Big)^2
+ (dz^4)^2+(dz^5)^2,\\
B &=& \frac{m}{1+m^2\Big[(z^4)^2+(z^5)^2\Big]}\Big(z^4dz^1\wedge dz^3+z^5dz^2\wedge dz^3\Big) \, .
\end{eqnarray}
This has T-duality monodromies in the $z^4$ and $z^5$ directions.

\subsection{$T^2$ bundle over $T^4$}\label{T2T4}
Consider the six-dimensional nilpotent Lie algebra whose  only non-vanishing commutators are
\begin{eqnarray}
[T_3,T_4] = mT_1,&\qquad& [T_3,T_5] = mT_2,\nonumber\\
{[T_5,T_6]} = mT_1,&\qquad& [T_4,T_6] = -mT_2.
\end{eqnarray}
Introducing coordinates $z^1,\dots z^6$,
the left-invariant one-forms are given by\begin{eqnarray}
P^1 &=& dz^1+m(z^4dz^3+z^6dz^5),\nonumber\\
P^2 &=& dz^2+m(z^5dz^3-z^6dz^4),\nonumber\\
P^3 &=& dz^3,\;
P^4 \;=\; dz^4,\nonumber\\
P^5 &=& dz^5,\;
P^6\;=\; dz^6
\end{eqnarray}
and the metric on the manifold is
\begin{equation}
ds^2 = \Big(dz^1+m(z^4dz^3+z^6dz^5)\Big)^2+\Big(dz^2+m(z^5dz^3-z^6dz^4)\Big)^2+(dz^3)^2+(dz^4)^2+(dz^5)^2+(dz^6)^2. \label{T2T4metric}
\end{equation}
This space  is a $T^2$ bundle over $T^4$
with fibre coordinates $z^1, z^2$.

T-duality in $z^1$ gives a $S^1$ bundle over $T^5$ with $H$-flux, given by
\begin{eqnarray}
ds^2 &=& (dz^1)^2+\Big(dz^2+m(z^5dz^3-z^6dz^4)\Big)^2+(dz^3)^2+(dz^4)^2+(dz^5)^2+(dz^6)^2,\\
H &=& -mdz^1\wedge dz^3 \wedge dz^4 - mdz^1\wedge dz^5 \wedge dz^6.
\end{eqnarray}
A further T-duality  in the $z^2$  direction gives a $T^6$ with $H$-flux, with a flat metric
\begin{eqnarray}
ds^2 = (dz^1)^2+(dz^2)^2+(dz^3)^2+(dz^4)^2+(dz^5)^2+(dz^6)^2,
\end{eqnarray}
and $H$-flux

\begin{eqnarray}
H &=&-mdz^1 \wedge dz^3 \wedge dz^4 -mdz^1 \wedge dz^5 \wedge dz^6\nonumber\\
& & -mdz^2 \wedge dz^3 \wedge dz^5 +mdz^2\wedge dz^4 \wedge dz^6  {.}
\end{eqnarray}
Starting from the metric (\ref{T2T4metric}) and   T-dualising in the $z^3$ direction gives a T-fold with metric and $B$-field
given by (\ref{T-fold1metric}) and (\ref{T-fold1Bfield}) in Appendix C.

\subsection{$T^3$ bundle over $T^3$}

The next case is  the six-dimensional nilpotent Lie algebra whose  only non-vanishing commutators are
\begin{eqnarray}
[T_5,T_6] = mT_1,&\qquad& [T_4,T_6] = -mT_2,\nonumber\\
{[T_4,T_5]} = mT_3.&\qquad&
\end{eqnarray}
 The left-invariant one-forms are given by
\begin{eqnarray}
P^1 &=& dz^1+mz^6dz^5,\nonumber\\
P^2 &=& dz^2-mz^6dz^4,\nonumber\\
P^3 &=& dz^3+mz^5dz^4,\nonumber\\
P^4 &=& dz^4,\;
P^5 \;=\; dz^5,\,
P^6\;=\; dz^6.
\end{eqnarray}
The metric is
\begin{equation}
ds^2 = \Big(dz^1+mz^6dz^5\Big)^2+\Big(dz^2-mz^6dz^4\Big)^2+\Big(dz^3+mz^5dz^4\Big)^2+(dz^4)^2+(dz^5)^2+(dz^6)^2.\label{T3T3metric}
\end{equation}
This is a $T^3$ bundle over $T^3$ with fibre coordinates $z^1,z^2,z^3$.

T-duality in the  $z^1$ direction gives a $T^2$ bundle over $T^4$ with $H$-flux. The metric and $H$-flux are
\begin{eqnarray}
ds^2 &=& (dz^1)^2+\Big(dz^2-mz^6dz^4\Big)^2+\Big(dz^3+mz^5dz^4\Big)^2+(dz^4)^2+(dz^5)^2+(dz^6)^2,\\
H &=& -mdz^1\wedge dz^5 \wedge z^6.
\end{eqnarray}
A further T-duality  in $z^2$ gives  an $S^1$ bundle over $T^5$ with $H$-flux. The metric and $H$-flux are
\begin{eqnarray}
ds^2 &=& (dz^1)^2+(dz^2)^2+\Big(dz^3+mz^5dz^4\Big)^2+(dz^4)^2+(dz^5)^2+(dz^6)^2,\\
H &= &-mdz^1\wedge dz^5 \wedge z^6 +mdz^2\wedge dz^4\wedge dz^6.
\end{eqnarray}
A final T-duality  in  the $z^3$ direction gives a $T^6$  with $H$-flux
\begin{eqnarray}
ds^2 = (dz^1)^2+(dz^2)^2+(dz^3)^2+(dz^4)^2+(dz^5)^2+(dz^6)^2,
\end{eqnarray}
\begin{equation}
H = -m dz^1\wedge dz^5 \wedge dz^6 - mdz^3 \wedge dz^4\wedge dz^5 +mdz^2\wedge dz^4 \wedge dz^6.
\end{equation}
 Starting with (\ref{T3T3metric}) and   T-dualising  in the  $z^4$ direction gives a T-fold with  metric and $B$-field given by
\begin{eqnarray}
ds^2 &=& (dz^1+mz^6dz^5)^2 +\frac{1}{1+m^2\Big[(z^5)^2+(z^6)^2\Big]}\Big[(dz)^2+(dz^3)^2+(dz^4)^2\Big]\nonumber\\
& & +\frac{1}{1+m^2\Big[(z^5)^2+(z^6)^2\Big]}\Big(mz^5dz^2+mz^6dz^3\Big)^2+(dz^5)^2+(dz^6)^2{,}\\
B &=& \frac{m}{1+m^2\Big[(z^5)^2+(z^6)^2\Big]}\Big(z^5 dz^3\wedge dz^4-z^6dz^2\wedge dz^4\Big) {.}
\end{eqnarray}

\subsection{$T^3$ bundle over $T^4$}\label{T3T4}

Next consider  the seven-dimensional nilpotent Lie algebra whose  only non-vanishing commutators are
\begin{eqnarray}
[T_4,T_5] = mT_1,&\qquad& [T_6,T_7] = mT_1,\nonumber\\
{[T_4,T_6]} = mT_2,&\qquad& [T_5,T_7] = -mT_2,\nonumber\\
{[T_4,T_7]} = mT_3,&\qquad& [T_5,T_6] = mT_3.
\end{eqnarray}
The left-invariant one-forms are given by
\begin{eqnarray}
P^1 &=& dz^1+m(z^5dz^4+z^7dz^6),\nonumber\\
P^2 &=& dz^2+m(z^6dz^4-z^7dz^5),\nonumber\\
P^3 &=& dz^3+m(z^7dz^4+z^6dz^5),\nonumber\\
P^4 &=& dz^4,\;
P^5 \;=\; dz^5,\nonumber\\
P^6&=& dz^6,\;
P^7\;=\; dz^7.
\end{eqnarray}
The metric is
\begin{eqnarray}
ds^2 &=& \Big(dz^1+m(z^5dz^4+z^7dz^6)\Big)^2+\Big(dz^2+m(z^6dz^4-z^7dz^5)\Big)^2+\Big(dz^3+m(z^7dz^4+z^6dz^5)\Big)^2\nonumber\\
& & +(dz^4)^2+(dz^5)^2+(dz^6)^2+(dz^7)^2.\label{T3T4metric}
\end{eqnarray}

T-duality in the  $z^1$ direction gives a nilmanifold with $H$-flux. The metric and $H$-flux are given
\begin{eqnarray}
ds^2 &=& (dz^1)^2+\Big(dz^2+m(z^6dz^4-z^7dz^5)\Big)^2+\Big(dz^3+m(z^7dz^4+z^6dz^5)\Big)^2\nonumber\\
& & +(dz^4)^2+(dz^5)^2+(dz^6)^2+(dz^7)^2,\\
H &=& -mdz^1\wedge dz^4\wedge dz^5-mdz^1\wedge dz^6\wedge dz^7.
\end{eqnarray}
A further T-duality   in the $z^2$ direction gives a nilmanifold with $H$-flux,  with metric and $H$-flux \begin{eqnarray}
ds^2 &=& (dz^1)^2+(dz^2)^2+\Big(dz^3+m(z^7dz^4+z^6dz^5)\Big)^2\nonumber\\
& & +(dz^4)^2+(dz^5)^2+(dz^6)^2+(dz^7)^2,\\
H &=& -mdz^1\wedge dz^4\wedge dz^5-mdz^1\wedge dz^6\wedge dz^7-mdz^2\wedge dz^4\wedge dz^6\\
&& +mdz^2\wedge dz^5\wedge dz^7.
\end{eqnarray}
 Next, T-duality   in the  $z^3$ direction gives a $T^7$ with $H$-flux. The metric and $H$-flux are given 
\begin{equation}
ds^2 = (dz^1)^2+(dz^2)^2+(dz^3)^2+(dz^4)^2+(dz^5)^2+(dz^6)^2+(dz^7)^2,
\end{equation}
\begin{eqnarray}
H =  &=& -mdz^1 \wedge dz^4\wedge dz^5 - mdz^1\wedge dz^6 \wedge dz^7 -mdz^2 \wedge dz^4\wedge dz^6 \nonumber\\
& &- mdz^3\wedge dz^4 \wedge dz^7 - mdz^3 \wedge dz^5 \wedge dz^6+mdz^2 \wedge dz^5\wedge dz^7.
\end{eqnarray}
Starting from the nilmanifold with metric (\ref{T3T4metric}) and   T-dualising in  the $z^4$ direction gives a T-fold with metric and $B$-field given by (\ref{T-fold2metric}), (\ref{T-fold2Bfield}) in Appendix C.

\subsection{$S^1$ bundle over $T^6$}

The last case is the seven-dimensional nilpotent Lie algebra whose only non-vanishing commutators are
\begin{eqnarray}
[T_2,T_3] = mT_1,&\qquad& [T_4,T_5] = mT_1,\nonumber\\
{[T_6,T_7]} = mT_1.&\qquad&
\end{eqnarray}
The left-invariant one-forms are given by
\begin{eqnarray}
P^1 &=& dz^1+m(z^3dz^2+z^5dz^4+z^7dz^6),\nonumber\\
P^2 &=& dz^2,\;
P^3 \;=\; dz^3,\nonumber\\
P^4 &=& dz^4,\;
P^5 \;=\; dz^5,\nonumber\\
P^6&=& dz^6,\;
P^7\;=\; dz^7.
\end{eqnarray}
The metric is
\begin{equation}
ds^2 = \Big(dz^1+m(z^3dz^2+z^5dz^4+z^7dz^6)\Big)^2+(dz^2)^2+(dz^3)^2+(dz^4)^2+(dz^5)^2+(dz^6)^2+(dz^7)^2.\label{T1T6metric}
\end{equation}
T-duality in the $z^1$ direction gives a $T^7$ with $H$-flux. The metric and $H$-flux are given
\begin{equation}
ds^2 = (dz^1)^2+(dz^2)^2+(dz^3)^2+(dz^4)^2+(dz^5)^2+(dz^6)^2+(dz^7)^2,
\end{equation}
\begin{equation}
H =  -mdz^1 \wedge dz^2\wedge dz^3 - mdz^1\wedge dz^4 \wedge dz^5 -mdz^1 \wedge dz^6\wedge dz^7.
\end{equation}

Starting with (\ref{T1T6metric}) and T-dualising in   $z^2$, $z^4$, and $z^6$ gives a T-fold with metric and $B$-field given by
\begin{eqnarray}
ds^2 &=& \frac{1}{1+m^2\Big[(z^3)^2+(z^5)^2+(z^7)^2\Big]}\Big[(dz^1)^2+(dz^2)^2+(dz^4)^2+(dz^6)^2\Big]\nonumber\\
& &+(dz^3)^2+(dz^5)^2+(dz^7)^2+\frac{1}{1+m^2\Big[(z^3)^2+(z^5)^2+(z^7)^2\Big]}\Big(mz^5dz^2-mz^3dz^4\Big)^2\nonumber\\
& &+\frac{1}{1+m^2\Big[(z^3)^2+(z^5)^2+(z^7)^2\Big]}\Big(mz^7dz^2-mz^3dz^6\Big)^2\nonumber\\
& &+\frac{1}{1+m^2\Big[(z^3)^2+(z^5)^2+(z^7)^2\Big]}\Big(mz^7dz^4-mz^5dz^6\Big)^2,
\end{eqnarray}
\begin{eqnarray}
B = \frac{m}{1+m^2\Big[(z^3)^2+(z^5)^2+(z^7)^2\Big]}\Big(z^3dz^1\wedge dz^2+z^5dz^1\wedge dz^4+z^7dz^1\wedge dz^6\Big).
\end{eqnarray}
This has T-duality monodromies in the $z^3,z^5$ and $z^7$ directions.


\section{The Nilfold Domain Wall Solution and its T-duals}

The nilfold ${\cal N}$ is not a consistent string background, but there are string solutions with nilfold fibres.
The simplest case is given by  ${\cal N}\times \mathbb{R}$ with 
\begin{equation}
ds^2 =V(\tau)(d\tau^2+dx^2+ dz^2 ) + \frac{1}{V(\tau)}(dy+mxdz)^2 \label {nilfold1}
\end{equation}
where
\begin{equation}
V(\tau) = m\tau+c
 \label{V}
\end{equation}
so that the metric is warped by a factor $V(\tau) $ depending on the coordinate $\tau$ on $\mathbb{R}$.
This is a hyperk\"ahler space which is singular at $\tau =-c/m$ where $V=0$.
Taking the product of this with 6-dimensional Minkowski space $\mathbb{R}^{1,5}$ provides a string background away from the singularity.

For a single domain wall,
 $V$ is piecewise linear
\begin{equation}
V(\tau) = \begin{cases}
c+m'\tau, & \tau \le 0 \\
c+m\tau , & \tau>0
\end{cases}\label{Vtau}
\end{equation}
with the solution (\ref{nilfold1}) for $\tau>0$ and the metric given by replacing $m$ with $m'$ for $ \tau \le 0$.
The function $V$ is continuous but not differentiable at $\tau =0$. The singularity at $\tau=0$ corresponds to a domain wall  at $\tau =0$ separating two \lq phases' with fluxes $m,m'$. This represents a brane that has a tension proportional to $m-m'$.
The product of this with 
6-dimensional Minkowski space 
 is    a dual  \cite{Hull:1998vy} of  the D8-brane solution \cite{Bergshoeff:1996ui}.
If $m'=-m$, then the solution can be identified under the reflection $\tau \to -\tau$ to give a single-sided domain wall \cite{Gibbons:1998ie}.

The multi-brane solution with  domain walls at  the points $\tau = \tau_1,\tau_2,\dots \tau_n$ is given by
\begin{equation}
V(\tau) = \begin{cases}
c_1+m_1\tau, & \tau \le \tau_1 \\
c_2+m_2\tau, & \tau_1 < \tau \le \tau_2\\
\vdots & \\
c_n+m_n\tau & \tau_{n-1} < \tau \le \tau_n\\
c_{n+1}+m_{n+1}\tau & \tau > \tau_n {,}
\end{cases}\label{Vtau1taun}
\end{equation}
for  constants $c_i$ and integers $ m_i$.
Continuity  fixes the constants $c_r$ for $r>1$   in terms of $c_1, m_i$ by 
\begin{equation}
c_{r+1}=c_r + (m_r-m_{r+1})\tau _r.
\end{equation}
The brane charge of the domain wall at $\tau _r$ is 
\begin{equation}
N_r=m_{r+1}- m_r .
\end{equation}
The derivative $M(\tau)\equiv V'(\tau)$ of $V(\tau)$ with respect to $\tau$  is piece-wise constant away from the domain wall points $\tau _r$
\begin{equation}
M(\tau) = \begin{cases}
m_1, & \tau < \tau_1 \\
m_2, & \tau_1 < \tau < \tau_2\\
\vdots & \\
m_n& \tau_{n-1} < \tau < \tau_n\\
m_{n+1}& \tau > \tau_n.
\end{cases}\label{M}
\end{equation}
The solution is then given by the metric
\begin{equation}
ds^2 =V(\tau)(d\tau^2+dx^2+ dz^2 ) + \frac{1}{V(\tau)}(dy+M(\tau)xdz)^2 \label {nilfold1}
\end{equation}
 and 
\begin{equation}
H=0, \qquad \Phi= \rm{constant}  {.} \label{nilfold2}
\end{equation}
 
 These hyperk\" ahler spaces are singular, but the singularities can be resolved as discussed in \cite{Gibbons:1998ie}, resulting in  non-singular hyperk\" ahler metrics given approximately by the metrics considered above. These multi-domain wall solutions are dual to multi D8 brane solutions. As D8 branes arise in consistent string theory configurations in type I$'$ string theory, these domain walls  arise 
 in a dual of type I$'$ theory, as discussed in \cite{Hull:1998vy}.
 
 T-dualising (\ref{nilfold1}) in the $y$ direction gives \cite{Hull:1998vy} the product of a $T^3$ with flux and a line, with
  the   conformally flat metric on $T^3\times \mathbb{R}$
\begin{equation}
ds^2 = V(\tau)(d\tau^2+dx^2+dy^2+dz^2)
\label{NS1}
\end{equation}
together with the   $H$-flux   
\begin{eqnarray}
H =  -M(\tau)dx\wedge dy\wedge dz \label{NS2}
\end{eqnarray}
and dilaton
\begin{equation}
e^{2\Phi} = V(\tau). \label{NS3}
\end{equation}

   T-duality in the $z$ direction gives  the T-fold fibred over a line \cite{Hull:2004in,Ellwood2006}. The metric and $B$-field of this background are given by 
\begin{equation}
ds^2 = V(\tau)(d\tau)^2+V(\tau)(dx)^2+\frac{V(\tau)}{V(\tau)^2+\Big(M(\tau)x\Big)^2}(dy^2+dz^2), 
\end{equation}
\begin{equation}
B = \frac{M(\tau)x}{V(\tau)^2+\Big(M(\tau)x\Big)^2} dy\wedge dz {,}
\end{equation}
while  the dilaton is
\begin{equation}
  \Phi=  \frac{1}{2}\log\left(\frac{V(\tau)}{V(\tau)^2+\Big(M(\tau)x\Big)^2}\right)
  \end{equation}


\section{Supersymmetric Domain Wall Solutions}

In the previous section, taking the product of the 3-dimensional nilfold with the real line  gave a space admitting a hyperk\" ahler metric.
Remarkably, a similar result applies for the nilmanifolds arising as  higher dimensional analogues of the nilfold of section \ref{Nilfolds} \cite{Gibbons:2001ds}.
Each of the spaces 
\begin{equation}
\mathcal{M} = \mathcal{G}/\Gamma 
\end{equation}
discussed in section \ref{Nilfolds}
is a $T^n$ bundle over $T^m$ for some $m,n$.
In each case, the space $\mathcal{M} \times \mathbb{R}$ admits a multi-domain wall type metric that has special holonomy \cite{Gibbons:2001ds}, so that taking the product of the domain wall solution
with Minkowski space gives a supersymmetric solution.
The solutions all  involve  a piecewise linear function $V(\tau) $ given by (\ref{Vtau1taun}) with derivative $M=V'$ given by (\ref{M}), corresponding to domain walls at the points $\tau_i$.
We now discuss each case in turn.


\subsection{6-dimensional domain wall solutions with $SU(3)$ holonomy}

6-dimensional solutions with  $SU(3)$ holonomy
can be constructed on $\mathcal{M} \times \mathbb{R}$
for the two cases of 5-dimensional nilfolds $\mathcal{M} $ discussed in section \ref{Nilfolds}, the
 $S^1$ bundle over $T^4$ and  the $T^2$ bundle over $T^3$.

\noindent \underline{Case 1: $S^1$ bundle over $T^4$}

The six-dimensional Calabi-Yau metric for this case is given by
\begin{eqnarray}
ds^2 &=& V^2(\tau)(d\tau)^2+V(\tau)\Big((dz^2)^2+(dz^3)^2+(dz^4)^2+(dz^5)^2\Big)\nonumber\\
& &+V^{-2}(\tau)\Big(dz^1+M(\tau)(z^3d z^2 +z^5 d z^4)\Big)^2,\label{metricS1T4}
\end{eqnarray}
where $\tau$ is a coordinate on the real line, $z^2, z^3, z^4$ and $z^5$ are coordinates on $T^4$, and $z^1$ is a coordinate on $S^1$.  The harmonic function $V(\tau)$ is given by (\ref{Vtau1taun}) and $M=V'$ (\ref{M}).
This metric is  K\"{a}hler Ricci-flat so it has $SU(3)$ holonomy and preserves $\frac{1}{4}$  supersymmetry. The K\"{a}hler form is given by
\begin{eqnarray}
J = d\tau \wedge \Big(dz^1+M(z^3 dz^2+z^5dz^4)\Big)-V(\tau)dz^2\wedge dz^3-V(\tau)dz^4\wedge dz^5.
\end{eqnarray}


\noindent\underline{Case 2: $T^2$ bundle over $T^3$}

The six-dimensional Calabi-Yau metric for this case is given by
\begin{eqnarray}
ds^2 &=& V^2(\tau)(d\tau)^2+V^2(\tau)(dz^3)^2+V(\tau)\Big((dz^4)^2+(dz^5)^2\Big)\nonumber\\
& &+V^{-1}(\tau)\Big(dz^1+Mz^4dz^3\Big)^2+V^{-1}(\tau)\Big(dz^2+Mz^5dz^3\Big)^2, \label{metricT2T3}
\end{eqnarray}
where $\tau$ is a coordinate on the real line, $z^3, z^4$ and $z^5$ are coordinates on $T^3$, while $z^1$ and $z^2$ are coordinates on $T^2$. The harmonic function $V(\tau)$ is given by (\ref{Vtau1taun}).
This metric is also K\"{a}hler Ricci-flat   with a K\"{a}hler form
\begin{equation}
J = V^2(\tau)d\tau\wedge dz^3+\Big(dz^1+Mz^4dz^3\Big)\wedge dz^4+\Big(dz^2+Mz^5dz^3\Big)\wedge dz^5 .
\end{equation}

\subsection{7-dimensional domain wall solutions with $G_2$ holonomy}

7-dimensional solutions with   $G_2$ holonomy
can be constructed on $\mathcal{M} \times \mathbb{R}$
for the two cases of 6-dimensional nilfolds $\mathcal{M}$ discussed in section \ref{Nilfolds}, the
 $T^2$ bundle over $T^4$ and  the $T^3$ bundle over $T^3$.
The $G_2$ holonomy implies the metrics are  Ricci-flat  and preserve $\frac{1}{8}$ of the supersymmetry.

\noindent\underline{Case 1: $T^2$ bundle over $T^4$}

The special holonomy metric in this case is given by
\begin{eqnarray}
ds^2 &=& V^4(\tau)(d\tau)^2+V^2(\tau)\Big((dz^3)^2+(dz^4)^2+(dz^5)^2+(dz^6)^2\Big)\nonumber\\
& &+V^{-2}(\tau)\Big(dz^1+M(z^4dz^3+z^6dz^5)\Big)^2\nonumber\\
& &+V^{-2}(\tau)\Big(dz^2+M(z^5dz^3-z^6dz^4)\Big)^2, \label{metricT2T4}
\end{eqnarray}
where $\tau$ is a coordinate on the real line, $z^3, z^4, z^5$ and  $z^6$ are coordinates on $T^4$, while $z^1$ and  $z^2$ are coordinates on $T^2$. 

\noindent\underline{Case 2: $T^3$ bundle over $T^3$}

The  metric in this case is given by
\begin{eqnarray}
ds^2 &=& V^3(\tau)(d\tau)^2+V^2(\tau)\Big((dz^4)^2+(dz^5)^2+(dz^6)^2\Big)+V^{-1}(\tau)\Big(dz^1+Mz^6dz^5\Big)^2\nonumber\\
& & V^{-1}(\tau)\Big(dz^2-Mz^6dz^4\Big)^2+V^{-1}(\tau)\Big(dz^3+Mz^5dz^4\Big)^2, \label{metricT3T3}
\end{eqnarray}
where $\tau$ is a coordinate on the real line, $z^4, z^5,$ and $z^6$ are coordinates on the $T^3$ base, while $z^1, z^2$ and  $z^3$ are coordinates on the $T^3$ fibre. The harmonic function $V(\tau)$ is given by (\ref{Vtau1taun}).

\subsection{8-dimensional domain wall solution with $Spin(7)$ holonomy}

In this case, an 8-dimensional solution with  $Spin(7)$ holonomy
can be constructed on $\mathcal{M} \times \mathbb{R}$ with $\mathcal{M}$ the 7-dimensional nilmanifold that is a $T^3$ bundle over $T^4$. The  metric is given by
\begin{eqnarray}
ds^2 &=& V^6(\tau)(d\tau)^2+V^3(\tau)\Big((dz^4)^2+(dz^5)^2+(dz^6)^2+(dz^7)^2)\Big)+V^{-2}(\tau)\Big(dz^1+M(z^5dz^4+z^7dz^6)\Big)^2\nonumber\\
& & +V^{-2}(\tau)\Big(dz^2+M(z^6dz^4-z^7dz^5)\Big)^2+V^{-2}(\tau)\Big(dz^3+M(z^7dz^4+z^6dz^5)\Big)^2, \label{metricT3T4}
\end{eqnarray}
where $\tau$ is a coordinate on the real line, $z^4, z^5, z^6$ and $z^7$ are coordinates on the $T^4$ base, and $z^1, z^2$ and  $z^3$ are coordinates on the $T^3$ fibre. The  function $V(\tau)$ is given by (\ref{Vtau1taun}).
The $Spin(7)$ holonomy 
implies the metric is    Ricci-flat   and preserves $\frac{1}{16}$ of the supersymmetry.

\subsection{8-dimensional domain wall solution with $SU(4)$ holonomy}
In this case, an 8-dimensional solution with $SU(4)$ holonomy can be constructed on $\mathcal{M} \times \mathbb{R}$ with $\mathcal{M}$ the 7-dimensional nilfold which is a $S^1$ bundle over $T^6$. 
The domain wall metric is given by
\begin{eqnarray}
ds^2 &=& V^3(\tau)(d\tau)^2+V(\tau)\Big((dz^2)^2+(dz^3)^2+(dz^4)^2+(dz^5)^2+(dz^6)^2+(dz^7)^2\Big)\nonumber\\
& & +V^{-3}(\tau)\Big(dz^1+M(z^3dz^2+z^5dz^4+z^7dz^6)\Big)^2,\label{metricS1T6}
\end{eqnarray}
where $\tau$ is a coordinate on the real line, $z^2, z^3, z^4, z^5, z^6$ and $z^7$ are coordinates on $T^6$ base, and $z^1$ is a coordinate on $S^1$ fibre. The  function $V(\tau)$ is given by (\ref{Vtau1taun}).
The $SU(4)$ holonomy 
implies the metrics is  Ricci-flat   and preserves $\frac{1}{8}$ of the supersymmetry.

\section{  
Special Holonomy Domain Walls    and Intersecting Branes  }

In this section, we will T-dualise each of the special holonomy domain wall solutions of the last section to obtain a system of intersecting branes. In each case, we obtain a standard intersecting brane configuration and check that it preserves exactly the same fraction of supersymmetry as the corresponding special holonomy   solution.

\subsection{$S^1$ fibred over $T^4$}\label{S1T4}
The supersymmetric Calabi-Yau solution has
ten-dimensional metric 
\begin{equation}
\label{flatmet}
ds_{10}^2 = ds^2(\mathbb{R}^{1,3})+ds_6^2,
\end{equation}
where $ds^2(\mathbb{R}^{1,n})$ is the flat metric of $(n+1)$-dimensional Minkowski space and 
 the 6-dimensional $SU(3)$ holonomy  metric is
\begin{eqnarray}
ds_6^2 &=& V^2(\tau)(d\tau)^2+V(\tau)\Big((dz^2)^2+(dz^3)^2+(dz^4)^2+(dz^5)^2\Big)\nonumber\\
& &+V^{-2}(\tau)\Big(dz^1+M(z^3d z^2 +z^5 d z^4)\Big)^2. \label{T1T4metricline}
\end{eqnarray}
The $H$-flux and the dilaton are trivial,
\begin{equation}
H = 0,\qquad \Phi = \text{constant}.
\end{equation}

T-duality in the $z^1$ direction gives the background with  metric \ref{flatmet} where
\begin{equation}
ds_6^2 = V^2(\tau)\Big((d\tau)^2+(dz^1)^2\Big)+V(\tau)\Big((dz^2)^2+(dz^3)^2+(dz^4)^2+(dz^5)^2\Big),
\end{equation}
 $H$-flux
\begin{equation}
H = -Mdz^1\wedge dz^2\wedge dz^3 - Mdz^1\wedge dz^4\wedge dz^5,
\end{equation}
and dilaton
\begin{equation}
e^{\Phi} = V(\tau).
\end{equation}
This solution describes two intersecting smeared NS5-branes.
\begin{center}
\begin{tabular}{|l|l|l|l|l|l|l|l|l|l|l|}
\hline
 & 0 & 1 & 2 & 3 & $z^1$ & $z^2$ & $z^3$ & $z^4$ & $z^5$ & $\tau$ \\ \hline
NS5 1 & $\times$ & $\times$ & $\times$ & $\times$ & $\bullet$ & $\bullet$ & $\bullet$ & $\times$ & $\times$ &  \\ \hline
NS5 2 & $\times$ & $\times$ & $\times$ & $\times$ & $\bullet$ & $\times$ & $\times$ & $\bullet$ & $\bullet$ &  \\ \hline
\end{tabular}
\end{center}
Here and in what follows,  $\times$ represents a world-volume direction and  $\bullet$ represents a smeared direction.
This then represents an NS5-brane lying in the $123z^4z^5$ directions and smeared over the $z^1z^2z^3$ directions which intersects an NS5-brane lying in the $123z^2z^3$ directions and smeared over the $z^1z^4z^4$ directions, with the intersection in the $123$ directions.
This intersection of two NS5-branes preserves $1/4$ supersymmetry \cite{Gauntlett:1998vk}.

S-duality give a background with intersecting D5-branes with metric 
\begin{equation}
ds_{10}^2 =V^{-1}(\tau)ds^2(\mathbb{R}^{1,3})+V(\tau)\Big((d\tau)^2+(dz^1)^2\Big)+\Big((dz^2)^2+(dz^3)^2+(dz^4)^2+(dz^5)^2\Big),
\end{equation}
 $RR$ 3-form field strength
\begin{equation}
F_{(3)} = -Mdz^1\wedge dz^2\wedge dz^3 - Mdz^1\wedge dz^4\wedge dz^5,
\end{equation}
and dilaton
\begin{equation}
e^{\Phi} = V^{-1}(\tau).
\end{equation}
This solution describes two intersecting D5-branes.
\begin{center}
\begin{tabular}{|l|l|l|l|l|l|l|l|l|l|l|}
\hline
 & 0 & 1 & 2 & 3 & $z^1$ & $z^2$ & $z^3$ & $z^4$ & $z^5$ & $\tau$ \\ \hline
D5 1 & $\times$ & $\times$ & $\times$ & $\times$ & $\bullet$ & $\bullet$ & $\bullet$ & $\times$ & $\times$ &  \\ \hline
D5 2 & $\times$ & $\times$ & $\times$ & $\times$ & $\bullet$ & $\times$ & $\times$ & $\bullet$ & $\bullet$ &  \\ \hline
\end{tabular}
\end{center}
 
T-duality in the  $z^1$, $z^2$ and $z^3$ directions gives a     D4-brane inside a D8-brane with the metric
\begin{equation}
ds_{10}^2 =V^{-1}(\tau)ds^2(\mathbb{R}^{1,3})+V(\tau)(d\tau)^2+V^{-1}(\tau)(dz^1)^2+\Big((dz^2)^2+(dz^3)^2+(dz^4)^2+(dz^5)^2\Big),
\end{equation}
and $RR$ fluxes
\begin{equation}
F_{(0)} = -M,\qquad
F_{(4)} = -Mdz^2\wedge dz^3\wedge dz^4 \wedge dz^5.
\end{equation}
and dilaton
\begin{equation}
e^{\Phi} = V^{-3/2}(\tau).
\end{equation}
This solution consists of a number of parallel D8-branes with a D4-brane inside each.
The D4-branes   each 
lie in the $123z^1$ directions and are smeared over the $z^2z^3z^4z^5$ directions.
\begin{center}
\begin{tabular}{|l|l|l|l|l|l|l|l|l|l|l|}
\hline
 & 0 & 1 & 2 & 3 & $z^1$ & $z^2$ & $z^3$ & $z^4$ & $z^5$ & $\tau$ \\ \hline
D8 & $\times$ & $\times$ & $\times$ & $\times$ & $\times$ & $\times$ & $\times$ & $\times$ & $\times$ &  \\ \hline
D4 & $\times$ & $\times$ & $\times$ & $\times$ & $\times$ & $\bullet$ & $\bullet$ & $\bullet$ & $\bullet$ &  \\ \hline
\end{tabular}
\end{center}
This is a standard example of a $1/4$ supersymmetric brane configuration and is T-dual to a D0-brane inside a D4-brane.

\subsection{$T^2$ fibred over $T^3$}
The $SU(3)$ holonomy metric in this case is  \begin{eqnarray}
ds_6^2 &=& V^2(\tau)(d\tau)^2+V^2(\tau)(dz^3)^2+V(\tau)\Big((dz^4)^2+(dz^5)^2\Big)\nonumber\\
& &+V^{-1}(\tau)\Big(dz^1+Mz^4dz^3\Big)^2+V^{-1}(\tau)\Big(dz^2+Mz^5dz^3\Big)^2.\label{T2T3metricline}
\end{eqnarray}
The ten-dimensional metric is 
\begin{equation}
ds_{10}^2 = ds^2(\mathbb{R}^{1,3})+ds_6^2.
\end{equation}
The $H$-flux and the dilaton are trivial,
\begin{equation}
H = 0,\qquad \Phi =  \text{constant}.
\end{equation}

T-duality in the $z^1$ and $z^2$ directions gives the metric
\begin{eqnarray}
ds_6^2 &=& V^2(\tau)\Big((d\tau)^2+(dz^3)^2\Big)\nonumber\\
& &+V(\tau)\Big((dz^1)^2+(dz^2)^2+(dz^4)^2+(dz^5)^2\Big)\nonumber
\end{eqnarray}
the $H$-flux
\begin{equation}
H = -Mdz^1\wedge dz^3\wedge dz^4 - Mdz^2\wedge dz^3\wedge dz^5,
\end{equation}
and the dilaton
\begin{equation}
e^{\Phi} = V(\tau).
\end{equation}
This solution represents two intersecting NS5-branes in the following configuration:
\begin{center}
\begin{tabular}{|l|l|l|l|l|l|l|l|l|l|l|}
\hline
 & 0 & 1 & 2 & 3 & $z^1$ & $z^2$ & $z^3$ & $z^4$ & $z^5$ & $\tau$ \\ \hline
NS5 1 & $\times$ & $\times$ & $\times$ & $\times$ & $\bullet$ & $\times$ & $\bullet$ & $\bullet$ & $\times$ &  \\ \hline
NS5 2 & $\times$ & $\times$ & $\times$ & $\times$ & $\times$ & $\bullet$ & $\bullet$ & $\times$ & $\bullet$ &  \\ \hline
\end{tabular}
\end{center}
S-duality takes this to a background with intersecting D5-branes with metric 
\begin{equation}
ds_{10}^2 = V^{-1}(\tau)ds^2(\mathbb{R}^{1,3})+V(\tau)\Big((d\tau)^2+(dz^3)^2\Big)+\Big((dz^1)^2+(dz^2)^2+(dz^4)^2+(dz^5)^2\Big),
\end{equation}
 $RR$ 3-form field strength
\begin{equation}
F_{(3)} = -Mdz^1\wedge dz^3\wedge dz^4 - Mdz^2\wedge dz^3\wedge dz^5,
\end{equation}
and dilaton
\begin{equation}
e^{\Phi} = V^{-1}(\tau).
\end{equation}
This solution represents two intersecting D5-branes:
\begin{center}
\begin{tabular}{|l|l|l|l|l|l|l|l|l|l|l|}
\hline
 & 0 & 1 & 2 & 3 & $z^1$ & $z^2$ & $z^3$ & $z^4$ & $z^5$ & $\tau$ \\ \hline
D5 1 & $\times$ & $\times$ & $\times$ & $\times$ & $\bullet$ & $\times$ & $\bullet$ & $\bullet$ & $\times$ &  \\ \hline
D5 2 & $\times$ & $\times$ & $\times$ & $\times$ & $\times$ & $\bullet$ & $\bullet$ & $\times$ & $\bullet$ &  \\ \hline
\end{tabular}
\end{center}
T-duality in the $z^1$, $z^3$ and $z^4$ directions gives a metric
\begin{equation}
ds_{10}^2 = V^{-1}(\tau)ds^2(\mathbb{R}^{1,3})+V(\tau)(d\tau)^2+V^{-1}(\tau)(dz^3)^2+\Big((dz^1)^2+(dz^2)^2+(dz^4)^2+(dz^5)^2\Big),
\end{equation}
 $RR$ fluxes
\begin{equation}
F_{(0)} = -M,\qquad
F_{(4)} = -Mdz^1\wedge dz^2\wedge dz^4 \wedge dz^5.
\end{equation}
and dilaton
\begin{equation}
e^{\Phi} = V^{-3/2}(\tau).
\end{equation}
This is the same solution as that found in section 6.1. 
and represents a D4-brane inside a D8-brane:
\begin{center}
\begin{tabular}{|l|l|l|l|l|l|l|l|l|l|l|}
\hline
 & 0 & 1 & 2 & 3 & $z^1$ & $z^2$ & $z^3$ & $z^4$ & $z^5$ & $\tau$ \\ \hline
D8 & $\times$ & $\times$ & $\times$ & $\times$ & $\times$ & $\times$ & $\times$ & $\times$ & $\times$ &  \\ \hline
D4 & $\times$ & $\times$ & $\times$ & $\times$ & $\bullet$ & $\bullet$ & $\times$ & $\bullet$ & $\bullet$ &  \\ \hline
\end{tabular}
\end{center}
This was to be expected as the $S^1$ over $T^4$ case is T-dual to the $T^2$ over $T^3$ case and gives the same solution as that found in section 6.1.

\subsection{$T^2$ fibred over $T^4$}
The domain wall metric in this case is given by
\begin{eqnarray}
ds_7^2 &=& V^4(\tau)(d\tau)^2+V^2(\tau)\Big(
(dz^3)^2+(dz^4)^2+(dz^5)^2+(dz^6)^2\Big)\nonumber\\
& &+V^{-2}(\tau)\Big(dz^1+M(z^4dz^3+z^6dz^5)\Big)^2\nonumber\\
& &+V^{-2}(\tau)\Big( dz^2+M(z^5dz^3-z^6dz^4)\Big)^2.
\end{eqnarray}
The ten-dimensional metric is 
\begin{equation}
ds_{10}^2 = ds^2(\mathbb{R}^{1,2})+ds_7^2.
\end{equation}
The $H$-flux and the dilaton are trivial,
\begin{equation}
H = 0,\qquad \Phi =  \text{constant}.
\end{equation}

T-duality in the $z^1$ and $z^2$ directions followed by the  coordinate transformations $z^4 \leftrightarrow z^5 $ and $z^5 \leftrightarrow z^6$
gives the metric 
\begin{eqnarray}
ds_7^2 = V^4(\tau)(d\tau)^2+V^2(\tau)\Big((dz^1)^2+(dz^2)^2+(dz^3)^2+(dz^4)^2+(dz^5)^2+(dz^6)^2\Big),
\end{eqnarray}
and $H$-flux
\begin{eqnarray}
H &=& -Mdz^1 \wedge dz^3 \wedge dz^6 -Mdz^1 \wedge dz^4 \wedge dz^5\nonumber\\
& & -Mdz^2 \wedge dz^3 \wedge dz^4 -Mdz^2\wedge dz^5 \wedge dz^6,
\end{eqnarray}
and dilaton
\begin{equation}
e^{\Phi} = V^2(\tau).
\end{equation}
This solution represents four  smeared NS5-branes intersecting in the 012 directions:
\begin{center}
\begin{tabular}{|l|l|l|l|l|l|l|l|l|l|l|}
\hline
 & 0 & 1 & 2 & $z^1$ & $z^2$ & $z^3$ & $z^4$ & $z^5$ & $z^6$ & $\tau$ \\ \hline
NS5 1 & $\times$ & $\times$ & $\times$ & $\bullet$ & $\times$ & $\bullet$ & $\times$ & $\times$ & $\bullet$ &  \\ \hline
NS5 2 & $\times$ & $\times$ & $\times$ & $\bullet$ & $\times$ & $\times$ & $\bullet$ & $\bullet$ & $\times$ &  \\ \hline
NS5 3 & $\times$ & $\times$ & $\times$ & $\times$ & $\bullet$ & $\bullet$ & $\bullet$ & $\times$ & $\times$ &  \\ \hline
NS5 4 & $\times$ & $\times$ & $\times$ & $\times$ & $\bullet$ & $\times$ & $\times$ & $\bullet$ & $\bullet$ &  \\ \hline
\end{tabular}
\end{center}
This intersection of four NS5-branes preserves $1/8$ supersymmetry \cite{Gauntlett:1998vk}.

S-duality give the metric
\begin{equation}
ds_{10}^2 = V^{-2}(\tau)ds^2(\mathbb{R}^{1,2})+V^2(\tau)(d\tau)^2+\Big((dz^1)^2+(dz^2)^2+(dz^3)^2+(dz^4)^2+(dz^5)^2+(dz^6)^2\Big),
\end{equation}
and $RR$ field strength
\begin{eqnarray}
F_{(3)} &=& -Mdz^1 \wedge dz^3 \wedge dz^6 -Mdz^1 \wedge dz^4 \wedge dz^5\nonumber\\
& & -Mdz^2 \wedge dz^3 \wedge dz^4 -Mdz^2\wedge dz^5 \wedge dz^6,
\end{eqnarray}
and dilaton
\begin{equation}
e^{\Phi} = V^{-2}(\tau).
\end{equation}
This solution describes four intersecting D5-branes.
\begin{center}
\begin{tabular}{|l|l|l|l|l|l|l|l|l|l|l|}
\hline
 & 0 & 1 & 2 & $z^1$ & $z^2$ & $z^3$ & $z^4$ & $z^5$ & $z^6$ & $\tau$ \\ \hline
 D5 1 & $\times$ & $\times$ & $\times$ & $\bullet$ & $\times$ & $\bullet$ & $\times$ & $\times$ & $\bullet$ &  \\ \hline
D5 2 & $\times$ & $\times$ & $\times$ & $\bullet$ & $\times$ & $\times$ & $\bullet$ & $\bullet$ & $\times$ &  \\ \hline
D5 3 & $\times$ & $\times$ & $\times$ & $\times$ & $\bullet$ & $\bullet$ & $\bullet$ & $\times$ & $\times$ &  \\ \hline
D5 4 & $\times$ & $\times$ & $\times$ & $\times$ & $\bullet$ & $\times$ & $\times$ & $\bullet$ & $\bullet$ &  \\ \hline
\end{tabular}
\end{center}
 
T-duality in the $z^1, z^3$ and $z^6$ directions gives the metric
\begin{equation}
ds_{10}^2 =V^{-2}(\tau)ds^2(\mathbb{R}^{1,2})+V^2(\tau)(d\tau)^2+\Big((dz^1)^2+(dz^2)^2+(dz^3)^2+(dz^4)^2+(dz^5)^2+(dz^6)^2\Big),
\end{equation}
and $RR$ field strength
\begin{eqnarray}
F_{(0)} &=& -M,\\
F_{(4)} &=& -Mdz^3 \wedge dz^4 \wedge dz^5 \wedge dz^6-Mdz^1 \wedge dz^2 \wedge dz^4\wedge dz^6\nonumber\\
& &-Mdz^1\wedge dz^2\wedge dz^3 \wedge dz^5  {,}
\end{eqnarray}
and dilaton
\begin{equation}
e^{\Phi} = V^{-2}(\tau).
\end{equation}
This solution represents 3 intersecting D4-branes inside   a D8-brane:
\begin{center}
\begin{tabular}{|l|l|l|l|l|l|l|l|l|l|l|}
\hline
 & 0 & 1 & 2 & $z^1$ & $z^2$ & $z^3$ & $z^4$ & $z^5$ & $z^6$ & $\tau$ \\ \hline
D8 & $\times$ & $\times$ & $\times$ & $\times$ & $\times$ & $\times$ & $\times$ & $\times$ & $\times$ &  \\ \hline
D4 1 & $\times$ & $\times$ & $\times$ & $\times$ & $\times$ & $\bullet$ & $\bullet$ & $\bullet$ & $\bullet$ &  \\ \hline
D4 2 & $\times$ & $\times$ & $\times$ & $\bullet$ & $\bullet$ &$\times$& $\bullet$ & $\times$ & $\bullet$ &  \\ \hline
D4 3 & $\times$ & $\times$ & $\times$ & $\bullet$ & $\bullet$ & $\bullet$ & $\times$ & $\bullet$ & $\times$ &  \\ \hline
\end{tabular}
\end{center}
T-dualising in the 1,2 directions and relabelling coordinates gives three mutually orthogonal D2-branes inside a D6-brane, with D2-branes in the 12, 34 and 56 planes all inside a D6-brane in the 123456 directions. This is a standard $1/8$ supersymmetric brane intersection.

\subsection{$T^3$ fibred over $T^3$}
The $G_2$ holonomy   metric in this case is given by
\begin{eqnarray}
ds_7^2 &=& V^3(\tau)(d\tau)^2+V^2(\tau)\Big((dz^4)^2+(dz^5)^2+(dz^6)^2\Big)+V^{-1}(\tau)\Big(dz^1+Mz^6dz^5\Big)^2\nonumber\\
& & +V^{-1}(\tau)\Big(dz^2-Mz^6dz^4\Big)^2+V^{-1}(\tau)\Big(dz^3+Mz^5dz^4\Big)^2. \label{T3T3metricline}
\end{eqnarray}
The ten-dimensional metric is 
\begin{equation}
ds_{10}^2 = ds^2(\mathbb{R}^{1,2})+ds_7^2.
\end{equation}
The $H$-flux and the dilaton are trivial,
\begin{equation}
H = 0,\qquad \Phi =  \text{constant}.
\end{equation}

T-duality in the $z^1$, $z^2$ and $z^3$ directions
followed by the coordinate transformation $z^2 \to -z^2$
 gives the metric
\begin{eqnarray}
ds_7^2 &=& V^3(\tau)(d\tau)^2+V(\tau)\Big((dz^1)^2+(dz^2)^2+(dz^3)^2\Big)\nonumber\\
& &+V^2(\tau)\Big((dz^4)^2+(dz^5)^2+(dz^6)^2\Big)  {,}
\end{eqnarray}
and $H$-flux
\begin{eqnarray}
H = -Mdz^1 \wedge dz^5\wedge dz^6 - Mdz^3\wedge dz^4 \wedge dz^5 -Mdz^2 \wedge dz^4\wedge dz^6,
\end{eqnarray}
and dilaton
\begin{equation}
e^{\Phi} = V^{3/2}(\tau).
\end{equation}
These solutions describe three intersecting smeared NS5-branes:
\begin{center}
\begin{tabular}{|l|l|l|l|l|l|l|l|l|l|l|}
\hline
 & 0 & 1 & 2 & $z^1$ & $z^2$ & $z^3$ & $z^4$ & $z^5$ & $z^6$ & $\tau$ \\ \hline
NS5 1 & $\times$ & $\times$ & $\times$ & $\bullet$ & $\times$ & $\times$ & $\times$ & $\bullet$ & $\bullet$ &  \\ \hline
NS5 2 & $\times$ & $\times$ & $\times$ & $\times$ & $\times$ & $\bullet$ & $\bullet$ & $\bullet$ & $\times$ &  \\ \hline
NS5 3 & $\times$ & $\times$ & $\times$ & $\times$ & $\bullet$ & $\times$ & $\bullet$ & $\times$ & $\bullet$ &  \\ \hline
\end{tabular}
\end{center}
This intersection of three NS5-branes preserves $1/8$ supersymmetry \cite{Gauntlett:1998vk}.

S-duality takes this to a solution with   metric
\begin{eqnarray}
ds_{10}^2 &=& V^{-3/2}(\tau)ds^2(\mathbb{R}^{1,2})+V^{3/2}(\tau)(d\tau)^2+V^{-1/2}(\tau)\Big((dz^1)^2+(dz^2)^2+(dz^3)^2\Big)\nonumber\\
& &+V^{1/2}(\tau)\Big((dz^4)^2+(dz^5)^2+(dz^6)^2\Big)  {,}
\end{eqnarray}
and $RR$ field strength
\begin{eqnarray}
F_{(3)} &=& -Mdz^1 \wedge dz^5\wedge dz^6 - Mdz^3\wedge dz^4 \wedge dz^5 -Mdz^2 \wedge dz^4\wedge dz^6,
\end{eqnarray}
and dilaton
\begin{equation}
e^{\Phi} = V^{-3/2}(\tau).
\end{equation}
This solution describes three intersecting D5-branes:
\begin{center}
\begin{tabular}{|l|l|l|l|l|l|l|l|l|l|l|}
\hline
 & 0 & 1 & 2 & $z^1$ & $z^2$ & $z^3$ & $z^4$ & $z^5$ & $z^6$ & $\tau$ \\ \hline
D5 1 & $\times$ & $\times$ & $\times$ & $\bullet$ & $\times$ & $\times$ & $\times$ & $\bullet$ & $\bullet$ &  \\ \hline
D5 2 & $\times$ & $\times$ & $\times$ & $\times$ & $\times$ & $\bullet$ & $\bullet$ & $\bullet$ & $\times$ &  \\ \hline
D5 3 & $\times$ & $\times$ & $\times$ & $\times$ & $\bullet$ & $\times$ & $\bullet$ & $\times$ & $\bullet$ &  \\ \hline
\end{tabular}
\end{center}
 
T-duality in the $z^1$, $z^5$ and $z^6$ directions  then gives the metric
\begin{eqnarray}
ds_{10}^2 &=&V^{-3/2}(\tau)ds^2(\mathbb{R}^{1,2})+V^{3/2}(\tau)(d\tau)^2+V^{-1/2}(\tau)\Big((dz^2)^2+(dz^3)^2 +(dz^5)^2+(dz^6)^2\Big)\nonumber\\
& &+V^{1/2}(\tau)\Big((dz^1)^2+(dz^4)^2\Big),
\end{eqnarray}
and $RR$ field strength
\begin{eqnarray}
F_{(0)} = -M,\qquad
F_{(4)} &=&- Mdz^1 \wedge dz^3\wedge dz^4 \wedge dz^6 -M dz^1 \wedge dz^2 \wedge dz^4\wedge dz^5,
\end{eqnarray}
and dilaton
\begin{equation}
e^{\Phi} = V^{-7/4}(\tau).
\end{equation}
This solution represents two intersecting D4-branes within a D8-brane:
\begin{center}
\begin{tabular}{|l|l|l|l|l|l|l|l|l|l|l|}
\hline
 & 0 & 1 & 2 & $z^1$ & $z^2$ & $z^3$ & $z^4$ & $z^5$ & $z^6$ & $\tau$ \\ \hline
D8 & $\times$ & $\times$ & $\times$ & $\times$ & $\times$ & $\times$ & $\times$ & $\times$ &$\times$ &  \\ \hline
D4 & $\times$ & $\times$ & $\times$ & $\bullet$& $\times$ & $\bullet$ & $\bullet$ & $\times$ & $\bullet$ &  \\ \hline
D4 & $\times$ & $\times$ & $\times$ & $\bullet$ & $\bullet$ & $\times$ & $\bullet$ &$\bullet$ & $\times$ &  \\ \hline
\end{tabular}
\end{center}
This is T-dual to two orthogonal D2-branes within a D6-brane, a standard $1/8$ supersymmetric brane configuration.
 
\subsection{$T^3$ fibred over $T^4$}
 The $Spin(7)$ holonomy metric is  given by
\begin{eqnarray}
ds_8^2 &=& V^6(\tau)(d\tau)^2+V^3(\tau)\Big((dz^4)^2+(dz^5)^2+(dz^6)^2+(dz^7)^2\Big)
\nonumber\\
& &
+V^{-2}(\tau)\Big(dz^1+M(z^5dz^4+z^7dz^6)\Big)^2  +V^{-2}(\tau)\Big(dz^2+M(z^6dz^4-z^7dz^5)\Big)^2
\nonumber\\
& &
 +V^{-2}(\tau)\Big(dz^3+M(z^7dz^4+z^6dz^5)\Big)^2.
\end{eqnarray}
The ten-dimensional metric is 
\begin{equation}
ds_{10}^2 = ds^2(\mathbb{R}^{1,1})+ds_8^2.
\end{equation}
The $H$-flux and the dilaton are trivial,
\begin{equation}
H = 0,\qquad  \Phi =  \text{constant}.
\end{equation}

T-duality in the $z^1$, $z^2$ and $z^3$ directions gives the metric
\begin{eqnarray}
ds_8^2 &=& V^6(\tau)(d\tau)^2+V^2\Big((dz^1)^2+(dz^2)^2+(dz^3)^2\Big)\nonumber\\
& &+V^3(\tau)\Big((dz^4)^2+(dz^5)^2+(dz^6)^2+(dz^7)^2\Big)  {,}
\end{eqnarray}
and $H$-flux
\begin{eqnarray}
H &=& -Mdz^1 \wedge dz^4\wedge dz^5 - Mdz^1\wedge dz^6 \wedge dz^7 -Mdz^2 \wedge dz^4\wedge dz^6 \nonumber\\
& &- Mdz^3\wedge dz^4 \wedge dz^7 - Mdz^3 \wedge dz^5 \wedge dz^6+Mdz^2 \wedge dz^5\wedge dz^7,
\end{eqnarray}
and dilaton
\begin{equation}
e^{\Phi} = V^{3}(\tau).
\end{equation}
This solution represents an intersection of five   NS5-branes and one anti-NS5-brane:
\begin{center}
\begin{tabular}{|l|l|l|l|l|l|l|l|l|l|l|}
\hline
 & 0 & 1 & $z^1$ & $z^2$ & $z^3$ & $z^4$ & $z^5$ & $z^6$ & $z^7$ & $\tau$ \\ \hline
NS5 1 & $\times$ & $\times$ & $\bullet$ & $\times$ & $\times$ & $\bullet$ & $\bullet$ & $\times$ & $\times$ &  \\ \hline
NS5 2 & $\times$ & $\times$ & $\bullet$ & $\times$ & $\times$ & $\times$ & $\times$ & $\bullet$ & $\bullet$ &  \\ \hline
NS5 3 & $\times$ & $\times$ & $\times$ & $\bullet$ & $\times$ & $\bullet$ & $\times$ & $\bullet$ & $\times$ &  \\ \hline
NS5 4 & $\times$ & $\times$ & $\times$ & $\times$ & $\bullet$ & $\bullet$ & $\times$ & $\times$ & $\bullet$ &  \\ \hline
NS5 5 & $\times$ & $\times$ & $\times$ & $\times$ & $\bullet$ & $\times$ & $\bullet$ & $\bullet$ & $\times$ &  \\ \hline
$\overline{\text{NS5}}$ & $\times$ & $\times$ & $\times$ & $\bullet$ & $\times$ & $\times$ & $\bullet$ & $\times$ & $\bullet$ &  \\ \hline
\end{tabular}
\end{center}
This intersection of six NS5-branes is one of the cases considered in \cite{Gauntlett:1998vk} and preserves $1/16$ supersymmetry .

S-duality then gives the metric
\begin{eqnarray}
ds_{10}^2 &=&V^{-3}(\tau)ds^2(\mathbb{R}^{1,1})+V^3(\tau)(d\tau)^2+V^{-1}(\tau)\Big((dz^1)^2+(dz^2)^2+(dz^3)^2\Big)\nonumber\\
& &+\Big((dz^4)^2+(dz^5)^2+(dz^6)^2+(dz^7)^2\Big)  {,}
\end{eqnarray}
and $RR$ field strength
\begin{eqnarray}
F_{(3)} &=& -Mdz^1 \wedge dz^4\wedge dz^5 - Mdz^1\wedge dz^6 \wedge dz^7 -Mdz^2 \wedge dz^4\wedge dz^6 \nonumber\\
& &- Mdz^3\wedge dz^4 \wedge dz^7 - Mdz^3 \wedge dz^5 \wedge dz^6+Mdz^2 \wedge dz^5\wedge dz^7,\end{eqnarray}
and dilaton
\begin{equation}
e^{\Phi} = V^{-3}(\tau) 
\end{equation}
changing the NS5-branes to D5-branes:
\begin{center}
\begin{tabular}{|l|l|l|l|l|l|l|l|l|l|l|}
\hline
 & 0 & 1 & $z^1$ & $z^2$ & $z^3$ & $z^4$ & $z^5$ & $z^6$ & $z^7$ & $\tau$ \\ \hline
D5 1 & $\times$ & $\times$ & $\bullet$ & $\times$ & $\times$ & $\bullet$ & $\bullet$ & $\times$ & $\times$ &  \\ \hline
D5 2 & $\times$ & $\times$ & $\bullet$ & $\times$ & $\times$ & $\times$ & $\times$ & $\bullet$ & $\bullet$ &  \\ \hline
D5 3 & $\times$ & $\times$ & $\times$ & $\bullet$ & $\times$ & $\bullet$ & $\times$ & $\bullet$ & $\times$ &  \\ \hline
D5 4 & $\times$ & $\times$ & $\times$ & $\times$ & $\bullet$ & $\bullet$ & $\times$ & $\times$ & $\bullet$ &  \\ \hline
D5 5 & $\times$ & $\times$ & $\times$ & $\times$ & $\bullet$ & $\times$ & $\bullet$ & $\bullet$ & $\times$ &  \\ \hline
$\overline{\text{D5}}$ & $\times$ & $\times$ & $\times$ & $\bullet$ & $\times$ & $\times$ & $\bullet$ & $\times$ & $\bullet$ &  \\ \hline
\end{tabular}
\end{center}
T-duality in the  $z^1, z^4$ and $z^5$ directions then gives the metric
\begin{eqnarray}
ds_{10}^2 &=& V^{-3}(\tau)(ds^2)(\mathbb{R}^{1,1})+V^3(\tau)(d\tau)^2+V(dz^1)^2+V^{-1}(\tau)\Big((dz^2)^2+(dz^3)^2\Big)\nonumber\\
& &+\Big((dz^4)^2+(dz^5)^2+(dz^6)^2+(dz^7)^2\Big) {,}
\end{eqnarray}
and RR field strength
\begin{eqnarray}
F_{(0)} &=& -M,\\
F_{(4)} &=&- Mdz^4 \wedge dz^5\wedge dz^6 \wedge dz^7- Mdz^1 \wedge dz^2\wedge dz^5 \wedge dz^6- Mdz^1 \wedge dz^3\wedge dz^5 \wedge dz^7\nonumber\\
& & - Mdz^1 \wedge dz^3\wedge dz^4 \wedge dz^6 +Mdz^1 \wedge dz^2\wedge dz^4 \wedge dz^7,
\end{eqnarray}
and dilaton
\begin{equation}
e^{\Phi} = V^{-5/2}(\tau).
\end{equation}
This is a $1/16$ supersymmetric configuration of four D4-branes and one anti-D4-brane intersecting inside a D8-brane:
\begin{center}
\begin{tabular}{|l|l|l|l|l|l|l|l|l|l|l|}
\hline
 & 0 & 1 & $z^1$ & $z^2$ & $z^3$ & $z^4$ & $z^5$ & $z^6$ & $z^7$ & $\tau$ \\ \hline
D8 & $\times$ & $\times$ & $\times$ & $\times$ & $\times$ & $\times$ & $\times$& $\times$ & $\times$ &  \\ \hline
D4 1 & $\times$ & $\times$ & $\times$ & $\times$ & $\times$ & $\bullet$ & $\bullet$ & $\bullet$ & $\bullet$ &  \\ \hline
D4 2 & $\times$ & $\times$ & $\bullet$& $\bullet$ & $\times$ &$\times$ & $\bullet$ & $\bullet$ & $\times$ &  \\ \hline
D4 3 & $\times$ & $\times$ & $\bullet$ &$\times$ & $\bullet$ &$\times$ & $\bullet$ & $\times$ & $\bullet$ &  \\ \hline
D4 4 & $\times$ & $\times$ &$\bullet$ & $\times$ & $\bullet$ & $\bullet$ & $\times$ & $\bullet$ & $\times$ &  \\ \hline
$\overline{\text{D4}}$ & $\times$ & $\times$ & $\bullet$ & $\bullet$ & $\times$ & $\bullet$ & $\times$& $\times$ & $\bullet$ &  \\ \hline
\end{tabular}
\end{center}

\subsection{$S^1$ fibred over $T^6$}
The $SU(4)$ holonomy metric is given by
\begin{eqnarray}
ds^2_8 &=& V^3(\tau)(d\tau)^2+V(\tau)\Big((dz^2)^2+(dz^3)^2+(dz^4)^2+(dz^5)^2+(dz^6)^2+(dz^7)^2\Big)\nonumber\\
& & +V^{-3}(\tau)\Big(dz^1+M(z^3dz^2+z^5dz^4+z^7dz^6)\Big)^2.\label{T1T6metricline}
\end{eqnarray}
The ten-dimensional metric is
\begin{equation}
ds^2_{10} = ds^2(\mathbb{R}^{1,1})+ds^2_8.
\end{equation}
The $H$-flux and the dilaton are trivial,
\begin{equation}
H = 0,\qquad  \Phi =  \text{constant}.
\end{equation}

T-duality in the $z^1$ direction gives the metric
\begin{equation}
ds^2_8 = V^3(\tau)\Big((d\tau)^2+(dz^1)^2\Big)+V(\tau)\Big((dz^2)^2+(dz^3)^2+(dz^4)^2+(dz^5)^2+(dz^6)^2+(dz^7)^2\Big)
\end{equation}
and $H$-flux
\begin{eqnarray}
H = -Mdz^1 \wedge dz^2\wedge dz^3 - Mdz^1\wedge dz^4 \wedge dz^5 -Mdz^1 \wedge dz^6\wedge dz^7,
\end{eqnarray}
and dilaton
\begin{equation}
e^{\Phi} = V^{3/2}(\tau).
\end{equation}
This solution describes three intersecting smeared NS5-branes:
\begin{center}
\begin{tabular}{|l|l|l|l|l|l|l|l|l|l|l|}
\hline
 & 0 & 1 & $z^1$ & $z^2$ & $z^3$ & $z^4$ & $z^5$ & $z^6$ & $z^7$ & $\tau$ \\ \hline
NS5 1 & $\times$ & $\times$ & $\bullet$ & $\bullet$ & $\bullet$ & $\times$ & $\times$ & $\times$ & $\times$ &  \\ \hline
NS5 2 & $\times$ & $\times$ & $\bullet$ & $\times$ & $\times$ & $\bullet$ & $\bullet$ & $\times$ & $\times$ &  \\ \hline
NS5 3 & $\times$ & $\times$ & $\bullet$ & $\times$ & $\times$ & $\times$ & $\times$ & $\bullet$ & $\bullet$ &  \\ \hline
\end{tabular}
\end{center}
This intersection of three NS5-branes preserves $1/8$ supersymmetry \cite{Gauntlett:1998vk}.

S-duality then takes this to the solution with metric
\begin{eqnarray}
ds_{10}^2 &=& V^{-3/2}(\tau)ds^2(\mathbb{R}^{1,1})+V^{3/2}(\tau)\Big((d\tau)^2+(dz^1)^2\Big)\nonumber\\
& &+V^{-1/2}(\tau)\Big((dz^2)^2+(dz^3)^2+(dz^4)^2+(dz^5)^2+(dz^6)^2+(dz^7)^2\Big)
\end{eqnarray}
and $RR$ field strength
\begin{eqnarray}
F_{(3)} &=& -Mdz^1 \wedge dz^2\wedge dz^3 - Mdz^1\wedge dz^4 \wedge dz^5 -Mdz^1 \wedge dz^6\wedge dz^7,
\end{eqnarray}
and dilaton
\begin{equation}
e^{\Phi} = V^{-3/2}(\tau),
\end{equation}
changing the NS5-branes to D5-branes
\begin{center}
\begin{tabular}{|l|l|l|l|l|l|l|l|l|l|l|}
\hline
 & 0 & 1 & $z^1$ & $z^2$ & $z^3$ & $z^4$ & $z^5$ & $z^6$ & $z^7$ & $\tau$ \\ \hline
D5 1 & $\times$ & $\times$ & $\bullet$ & $\bullet$ & $\bullet$ & $\times$ & $\times$ & $\times$ & $\times$ &  \\ \hline
D5 2 & $\times$ & $\times$ & $\bullet$ & $\times$ & $\times$ & $\bullet$ & $\bullet$ & $\times$ & $\times$ &  \\ \hline
D5 3 & $\times$ & $\times$ & $\bullet$ & $\times$ & $\times$ & $\times$ & $\times$ & $\bullet$ & $\bullet$ &  \\ \hline
\end{tabular}
\end{center}

T-duality in the $z^1$, $z^2$ and  $z^3$ directions then gives the metric
\begin{eqnarray}
ds_{10}^2 &=& V^{-3/2}(\tau)ds^2(\mathbb{R}^{1,1})+V^{3/2}(\tau)(d\tau)^2+V^{-3/2}(dz^1)^2\nonumber\\
& &+V^{1/2}\Big((dz^2)^2+(dz^3)^2\Big)+V^{-1/2}(\tau)\Big((dz^4)^2+(dz^5)^2+(dz^6)^2+(dz^7)^2\Big),
\end{eqnarray}
 $RR$ field strength
\begin{eqnarray}
F_{(0)} &=& -M\\
F_{(4)} &=&- Mdz^2 \wedge dz^3\wedge dz^4 \wedge dz^5- Mdz^2 \wedge dz^3\wedge dz^6 \wedge dz^7,
\end{eqnarray}
and dilaton
\begin{equation}
e^{\Phi} = V^{-7/4}(\tau).
\end{equation}
This solution represents two intersecting D4-branes within a D8-brane:
\begin{center}
\begin{tabular}{|l|l|l|l|l|l|l|l|l|l|l|}
\hline
 & 0 & 1 & $z^1$ & $z^2$ & $z^3$ & $z^4$ & $z^5$ & $z^6$ & $z^7$ & $\tau$ \\ \hline
D8 & $\times$ & $\times$ & $\times$ & $\times$ & $\times$ & $\times$ & $\times$ & $\times$ &$\times$ &  \\ \hline
D4 & $\times$ & $\times$ & $\times$ & $\bullet$& $\bullet$& $\bullet$ & $\bullet$ & $\times$ & $\times$ &  \\ \hline
D4 & $\times$ & $\times$ & $\times$ & $\bullet$ & $\bullet$ & $\times$ & $\times$ &$\bullet$ & $\bullet$ &  \\ \hline
\end{tabular}
\end{center}



\section{T-folds fibred over line}
\subsection{T-fold from the  $S^1$ bundle over $T^4$ fibred over a line}
Starting from the metric (\ref{T1T4metricline}) and   T-dualising along the $z^2$ and $z^4$ directions gives a T-fold background with the metric and $B$-field
\begin{eqnarray}
ds_6^2 &=& V^2(\tau) (d\tau)^2 +\frac{V(\tau)}{V^3(\tau)+M^2\Big[(z^3)^2+(z^5)^2\Big]}(dz^1)^2+\frac{V^2(\tau)}{V^3(\tau)+M^2\Big[(z^3)^2+(z^5)^2\Big]}(dz^2)^2\nonumber\\
& &+\frac{M^2}{V(\tau)\Big(V^3(\tau)+M^2\Big[(z^3)^2+(z^5)^2\Big]\Big)}(z^5dz^2-z^3dz^4)^2+\frac{V^2(\tau)}{V^3(\tau)+M^2\Big[(z^3)^2+(z^5)^2\Big]}(dz^4)^2\nonumber\\
& &+V(\tau)[(dz^3)^2+(dz^5)^2],\label{T-foldmetricS1T4}
\end{eqnarray}
\begin{equation}
B = \frac{M}{V^3(\tau)+M^2\Big[(z^3)^2+(z^5)^2\Big]}(z^3dz^1\wedge dz^2+z^5dz^1\wedge dz^4)
\end{equation}
and the dilaton
\begin{equation}
e^{2\Phi} = \frac{V(\tau)}{V^3(\tau)+M^2[(z^3)^2+(z^5)^2]}.
\end{equation}
Since the $S^1$ bundle over $T^4$ is T-dual to the $T^2$ bundle over $T^3$, doing T-duality in the $z^3$ direction of the metric (\ref{T2T3metricline}) will result in the same T-fold. 
\subsection{T-fold from the $T^3$ bundle over $T^3$ fibred over a line}

Starting from the metric (\ref{T3T3metricline}) and   T-dualising in the $z^4$ direction gives a T-fold background with   metric and $B$-field
\begin{eqnarray}
ds^2 &=& V^3(\tau)(d\tau)^2+\frac{1}{V(\tau)}(dz^1+Mz^6dz^5)^2 +\frac{V^2(\tau)}{V^3(\tau)+M^2\Big[(z^5)^2+(z^6)^2\Big]}[(dz^2)^2+(dz^3)^2]+\nonumber\\
& & \frac{V(\tau)}{V^3(\tau)+M^2\Big[(z^5)^2+(z^6)^2\Big]}(dz^4)^2+\frac{M^2}{V(\tau)\Big({V^3(\tau)+M^2\Big[(z^5)^2+(z^6)^2\Big]}\Big)}(z^5dz^2+z^6dz^3)^2\nonumber\\& &+V^2(\tau)(dz^5)^2+V^2(\tau)(dz^6)^2,\\
B &=& \frac{M}{V^3(\tau)+M^2\Big[(z^5)^2+(z^6)^2\Big]}(z^5 dz^3\wedge dz^4-z^6dz^2\wedge dz^4)
\end{eqnarray}
and   dilaton
\begin{equation}
e^{2\Phi} = \frac{V(\tau)}{V^3(\tau)+M^2\Big[(z^5)^2+(z^6)^2\Big]}.
\end{equation}


\subsection{T-fold from the $S^1$ bundle over $T^3$ fibred over a line}
Starting from the metric (\ref{T1T6metricline}) and   T-dualising in the $z^2, z^4$ and $z^6$ directions gives a T-fold background with   metric and $B$-field
\begin{eqnarray}
ds^2 &=& V^3(\tau)(d\tau^2)+V(\tau)\Big((dz^3)^2+(dz^5)^2+(dz^7)^2\Big)+ \frac{V(\tau)}{V^4(\tau)+M^2\Big[(z^3)^2+(z^5)^2+(z^7)^2\Big]}(dz^1)^2\nonumber\\
& &+\frac{V^3(\tau)}{V^4(\tau)+M^2\Big[(z^3)^2+(z^5)^2+(z^7)^2\Big]}\Big[(dz^2)^2+(dz^4)^2+(dz^6)^2\Big]\nonumber\\
& &+\frac{M^2}{V(\tau)\Big(V^4(\tau)+M^2\Big[(z^3)^2+(z^5)^2+(z^7)^2\Big]\Big)}\Big(z^5dz^2-z^3dz^4\Big)^2\nonumber\\
& &+\frac{M^2}{V(\tau)\Big(V^4(\tau)+M^2\Big[(z^3)^2+(z^5)^2+(z^7)^2\Big]\Big)}\Big(z^7dz^2-z^3dz^6\Big)^2\nonumber\\
& &+\frac{M^2}{V(\tau)\Big(V^4(\tau)+M^2\Big[(z^3)^2+(z^5)^2+(z^7)^2\Big]\Big)}\Big(z^7dz^4-z^5dz^6\Big)^2
,\\
B &=&\frac{M}{V^4(\tau)+M^2\Big[(z^3)^2+(z^5)^2+(z^7)^2\Big]}\Big(z^3dz^1\wedge dz^2+z^5dz^1\wedge dz^4+z^7dz^1\wedge dz^6\Big)
\end{eqnarray}
and   dilaton
\begin{equation}
e^{2\Phi} = \frac{V(\tau)}{V^4(\tau)+M^2\Big[(z^3)^2+(z^5)^2+(z^7)^2\Big]}
\end{equation}

\subsection{Further T-folds}

The T-folds arising  from T-dualising  the $T^2$ bundle over $T^4$ fibred    over a line of section \ref{T2T4} and  from T-dualising  the $T^3$ bundle over $T^4$ fibred    over a line of section \ref{T3T4} 
have complicated forms for the metric and B-field and are given in Appendix C.



\section{Special Holonomy Metrics Specified by Multiple Functions}

Our starting point has been the special holonomy metrics of \cite{Gibbons:2001ds} specified by a single piecewise linear function $V$.
Dualising gave   intersecting brane solutions specified by a single function.
However, the intersecting brane solutions can be generalised by introducing a separate   linear function for each brane while preserving the same amount of supersymmetry. Dualising back then gives a generalisation of the original  special holonomy metric involving a number of  functions instead of just one. This in turn generalises the nilmanifold
fibres to more general torus bundles than those considered previously. This leads to solutions specified by piecewise linear functions $V_1 (\tau) ,V_2  (\tau),\dots$.
The simplest case is to take all of these linear:
\begin{equation}
V_r(\tau) = m^{(r)}(\tau - \tau^{(r)})+c ^{(r)} , \qquad r=1,2,\dots
 \label{Vr}
\end{equation}
with different choices for the locations $\tau^{(r)}$, slopes $m^{(r)}$ and constants $c^{(r)}$ for each function.
More generally, they can each be taken to be piecewise linear:
\begin{equation}
V_r(\tau) = \begin{cases}
c_1^{(r)}+m_1^{(r)}\tau, & \tau \le \tau_1^{(r)} \\
c_2^{(r)}+m_2^{(r)}\tau, & \tau_1^{(r)} < \tau \le \tau_2^{(r)}\\
\vdots & \\
c_n^{(r)}+m_n^{(r)}\tau {,} & \tau_{n-1}^{(r)} < \tau \le \tau_n^{(r)}\\
c_{n+1}^{(r)}+m_{n+1}^{(r)}\tau {,} & \tau > \tau_n^{(r)}
\end{cases}\label{Vtau1taunr}
\end{equation}
with the continuity condition
\begin{equation}
c_{i+1}^{(r)} =c_i ^{(r)} + (m_i^{(r)} -m_{i+1}^{(r)} )\tau _i ^{(r)} .
\end{equation}
Then the derivatives $M_r(\tau)\equiv V_r'(\tau)$ are
\begin{equation}
M_r(\tau) = \begin{cases}
m_1 ^{(r)}, & \tau < \tau_1 ^{(r)} \\
m_2 ^{(r)}, & \tau_1 ^{(r)} < \tau < \tau_2 ^{(r)}\\
\vdots & \\
m_n ^{(r)} {,}& \tau_{n-1} ^{(r)} < \tau < \tau_n ^{(r)}\\
m_{n+1} ^{(r)} {,}& \tau > \tau_n ^{(r)}.
\end{cases}\label{Mr}
\end{equation}

We now give the generalisation of each of the special holonomy metrics to ones with multiple functions.
Dualising will generalisations of the give T-folds given in section 7 that are specified by multiple functions.

\subsection{$S^1$ bundle over $T^4$}

The $SU(3)$ holonomy metric (\ref{metricS1T4}) specified by a single function $V$ was dualised in subsection \ref{S1T4} to a solution with two intersecting D5-branes, specified by a single function.
This can be generalised by going to the general solution for two  intersecting D5-branes with 
 two different functions $V_1$, $V_2$ corresponding to the two D5-branes. Dualising back then gives an $SU(3)$ holonomy metric
   given by
\begin{eqnarray}
ds^2 &=& V_1(\tau)V_2(\tau)d\tau^2+\frac{1}{V_1(\tau)V_2(\tau)}\Big(dz^1+M_1(\tau)z^3dz^2+M_2(\tau)z^5dz^4\Big)^2\nonumber\\
& &+V_1(\tau)\Big((dz^2)^2+(dz^3)^2\Big)+V_2(\tau)\Big((dz^4)^2+(dz^5)^2\Big).
\end{eqnarray}
In the case in which $V_1,V_2$ are linear, given by (\ref{Vr}), 
 the nilmanifold fibre is an $S^1$ bundle over $T^4$ with 
 first Chern class  given by
\begin{eqnarray}
c = -m^{(1)}dz^2\wedge dz^3 -m^{(2)}dz^4\wedge dz^5
\end{eqnarray}
specified by the two integers; $m^{(1)}$ and  $m^{(2)}$. 
Taking $V_1=V_2$ so that $m^{(1)}=m^{(2)}$ recovers the solution (\ref{metricS1T4}).
More generally, both $V_1$ and $V_2$ can be taken to piecewise linear, of the form (\ref{Vtau1taunr}), and there are different values of the integers; $m^{(1)}$ and  $m^{(2)}$ for different values of $\tau$.

\subsection{$T^2$ bundle over $T^3$}

The  $SU(3)$ holonomy solution (\ref{metricT2T3}) is  dual to the same solution with two intersecting D5-branes as the previous case, and so again has a generalisation with two linear functions.
The $SU(3)$ holonomy metric is given by
\begin{eqnarray}
ds^2 &=& V_1(\tau)V_2(\tau)d\tau^2+\frac{1}{V_1(\tau)}\Big(dz^1+M_1z^4dz^3\Big)^2+\frac{1}{V_2(\tau)}\Big(dz^2+M_2z^5dz^3\Big)^2\nonumber\\
& & +V_1(\tau)V_2(\tau)(dz^3)^2+V_1(\tau)(dz^4)^2+V_2(\tau)(dz^5)^2.
\end{eqnarray}
For linear functions of the form (\ref{Vr}),
the nilmanifold fibre is a $T^2$ bundle over $T^3$ with   
 two first Chern classes 
 \begin{eqnarray}
c_1 &=& -m^{(1)}dz^3\wedge dz^4,\\
c_2 &=& -m^{(2)} dz^3 \wedge dz^5.
\end{eqnarray}

\subsection{$T^2$ bundle over $T^4$}

The $G_2$ holonomy solution (\ref{metricT2T4}) is dual to a solution with four intersecting branes, and so has a generalisation with four independent 
piecewise linear functions $V_1(\tau), V_2(\tau), V_3(\tau),$ and $V_4(\tau)$ to give the $G_2$ holonomy metric 
\begin{eqnarray}
ds^2 &=& V_1(\tau)V_2(\tau)V_3(\tau)V_4(\tau)d\tau^2+\frac{1}{V_1(\tau)V_2(\tau)}\Big(dz^1+M_1 z^4dz^3+M_2 z^6dz^5\Big)^2\nonumber\\
& & +\frac{1}{V_3(\tau)V_4(\tau)}\Big(dz^2+M_3z^5dz^3-M_4z^6dz^4\Big)^2+V_1(\tau)V_3(\tau)(dz^3)^2\nonumber\\
& &+ V_1(\tau)V_4(\tau)(dz^4)^2+V_2(\tau)V_3(\tau)(dz^5)^2+V_2(\tau)V_4(\tau)(dz^6)^2.
\end{eqnarray}
For linear functions of the form (\ref{Vr}),
the nilmanifold fibre is a $T^2$ bundle over $T^4$ with
 two first Chern classes 
\begin{eqnarray}
c_1 = -m^{(1)}dz^3 \wedge dz^4-m^{(2)} dz^5 \wedge dz^6,
c_2 = -m^{(3)} dz^3 \wedge dz^5 + m^{(4)}dz^4\wedge dz^6.
\end{eqnarray}
In total, there are 4 numbers, $m^{(r)}$, $r=1,2,3,4$ that determine this space.

\subsection{$T^3$ bundle over $T^3$}

The $G_2$ holonomy solution (\ref{metricT3T3}) is dual to a solution with three intersecting branes, and so has a generalisation with three independent 
piecewise linear functions $V_1(\tau), V_2(\tau),$ and $V_3(\tau)$ to give the $G_2$ holonomy metric 
\begin{eqnarray}
ds^2 &=& V_1(\tau)V_2(\tau)V_3(\tau)d\tau^2+\frac{1}{V_1}\Big(dz^1+M_1z^6dz^5\Big)^2+\frac{1}{V_2(\tau)}\Big(dz^2-M_2z^5dz^4\Big)^2\nonumber\\
& & +\frac{1}{V_3(\tau)}\Big(dz^3+M_3z^5dz^4\Big)^2+V_2(\tau)V_3(\tau)(dz^4)^2+V_1(\tau)V_3(\tau)(dz^5)^2\nonumber\\
& &+V_1(\tau)V_2(\tau)(dz^6)^2.
\end{eqnarray}
If the functions are of the form  (\ref{Vr}), then
the nilmanifold is a  $T^3$ bundle over $T^3$
with three first Chern classes, $c_1, c_2$ and  $c_3$ \begin{eqnarray}
c_1 &=& -m^{(1)}dz^5 \wedge dz^6\nonumber\\
c_2 &=& m^{(2)}dz^4\wedge dz^6 \nonumber\\
c_3 &=&-m^{(3)} dz^4 \wedge dz^5.
\end{eqnarray}

\subsection{$T^3$ bundle over $T^4$}

The $Spin(7)$ holonomy metric (\ref{metricT3T4}) is dual to a solution with six intersecting branes and so has a generalisation to a $Spin(7)$ holonomy metric specified by six functions $V_1(\tau),\cdots, V_6(\tau)$, with 
metric \begin{eqnarray}
ds^2 &=& V_1(\tau)V_2(\tau)V_3(\tau)V_4(\tau)V_5(\tau)V_6(\tau)d\tau^6+\frac{1}{V_1(\tau)V_2(\tau)}\Big(dz^1+M_1z^5dz^4+M_2z^7dz^6\Big)^2\nonumber\\
& &+\frac{1}{V_3(\tau)V_4(\tau)}\Big(dz^2+M_3z^6dz^4-M_4z^7dz^5\Big)^2+\frac{1}{V_5(\tau)V_6(\tau)}\Big(dz^3+M_5z^7dz^4+M_6z^6dz^5\Big)^2\nonumber\\
& &+V_1(\tau)V_3(\tau)V_5(\tau)(dz^4)^2+V_1(\tau)V_4(\tau)V_6(\tau)(dz^5)^2\nonumber\\
& &+V_2(\tau)V_3(\tau)V_6(\tau)(dz^6)^2+V_2(\tau)V_4(\tau)V_5(\tau)(dz^7)^2.
\end{eqnarray}
If the functions are of the form (\ref{Vr}), then
the nilmanifold is a $T^3$ bundle over $T^4$
with three first Chern classes,
\begin{eqnarray}
c_1 &=& -m^{(1)}dz^4 \wedge dz^5-m^{(2)}dz^6\wedge dz^7\nonumber\\
c_2 &=& -m^{(3)}dz^4 \wedge dz^6+m^{(4)}dz^5\wedge dz^7 \nonumber\\
c_3 &=&-m^{(5)}dz^4 \wedge dz^7-m^{(6)}dz^5\wedge dz^6.
\end{eqnarray}
There are 6 numbers, $m^{(r)}$, $r=1,\cdots, 6$ that determine this space.

\subsection{$S^1$ bundle over $T^6$}

The $SU(4)$ holonomy solution (\ref{metricS1T6}) is dual to a solution with three intersecting branes, and so has a generalisation with three independent 
piecewise linear functions $V_1(\tau), V_2(\tau),$ and $V_3(\tau)$ to give the $SU(4)$ holonomy metric 
\begin{eqnarray}
ds^2 &=& V_1(\tau)V_2(\tau)V_3(\tau)d\tau^2+\frac{1}{V_1(\tau)V_2(\tau)V_3(\tau)}\Big(dz^1+M_1z^3dz^2+M_2z^5dz^4+M_3z^7dz^6\Big)^2\nonumber\\
& &+V_1(\tau)\Big((dz^2)^2+(dz^3)^2\Big)+V_2(\tau)\Big((dz^4)^2+(dz^5)^2\Big)\nonumber\\
&&+V_3(\tau)\Big((dz^6)^2+(dz^7)^2\Big)  {.}
\end{eqnarray}
If the functions are of the form  (\ref{Vr}), then
the fibre is a $T^3$ bundle over $T^6$
with  first Chern class is
\begin{eqnarray}
c = -m^{(1)}dz^2\wedge dz^3 -m^{(2)}dz^4\wedge dz^5 -m^{(3)}dz^6\wedge dz^7.
\end{eqnarray}


\section*{Acknowledgments}

We are grateful to Amihay Hanany, Bobby Acharya, Jerome Gauntlett  and Simon Salamon for helpful discussions. 
 The  work of CH is supported by the EPSRC Programme
Grant EP/K034456/1,
and by the STFC
Consolidated Grant ST/L00044X/1.


\appendix
\section{Appendix: Calculation of left-invariant 1-forms}
In this and the following appendix, we give some of the details of the calculations leading to the results in the body of the paper for the case of the $S^1$ bundle over $T^4$  introduced in section 3.1. The other cases are treated similarly.

Let $\mathcal{G}_5$ be the five dimensional nilpotent Lie group with non-vanishing commutators
\begin{equation}
[T_2,T_3] = mT_1, \qquad [T_4,T_5] = mT_1.
\end{equation}
This is the algebra arising from the $S^1$ bundle over $T^4$ in 3.1.
A group element $g$ (in a neighbourhood of the identity) can be written as
\begin{equation}
g = \exp(z^1T_1)\exp(z^2T_2)\exp(z^3T_3)\exp(z^4T_4)\exp(z^5T_5), \label{gelement}
\end{equation}
where $z^1,\cdots, z^5$ are local coordinates on $\mathcal{G}_5.$
The inverse of $g$ is
\begin{equation}
g^{-1} = \exp(-z^5T_5)\exp(-z^4T_4)\exp(-z^3T_3)\exp(-z^2T_2)\exp(-z^1T_1).
\end{equation}
From (\ref{gelement}), $dg$ is given by
\begin{eqnarray}
dg &=& [dz^1T_1]\exp(z^1T_1)\exp(z^2T_2)\exp(z^3T_3)\exp(z^4T_4)\exp(z^5T_5)\nonumber\\
& & +\exp(z^1T_1)[dz^2T_2]\exp(z^2T_2)\exp(z^3T_3)\exp(z^4T_4)\exp(z^5T_5)\nonumber\\
& & +\exp(z^1T_1)\exp(z^2T_2)[dz^3T_3]\exp(z^3T_3)\exp(z^4T_4)\exp(z^5T_5)\nonumber\\
& & +\exp(z^1T_1)\exp(z^2T_2)\exp(z^3T_3)[dz^4T_4]\exp(z^4T_4)\exp(z^5T_5)\nonumber\\
& &+\exp(z^1T_1)\exp(z^2T_2)\exp(z^3T_3)\exp(z^4T_4)[dz^5T_5]\exp(z^5T_5).
\end{eqnarray}
The left-invariant one-form is then given by
\begin{eqnarray}
g^{-1}dg = P^aT_a.
\end{eqnarray}
\begin{eqnarray}
g^{-1}dg &=& \exp(-z^5T_5)\exp(-z^4T_4)\exp(-z^3T_3)\exp(-z^2T_2)\exp(-z^1T_1)\times\nonumber\\
& & \Big([dz^1T_1]\exp(z^1T_1)\exp(z^2T_2)\exp(z^3T_3)\exp(z^4T_4)\exp(z^5T_5)\nonumber\\
& & +\exp(z^1T_1)[dz^2T_2]\exp(z^2T_2)\exp(z^3T_3)\exp(z^4T_4)\exp(z^5T_5)\nonumber\\
& & +\exp(z^1T_1)\exp(z^2T_2)[dz^3T_3]\exp(z^3T_3)\exp(z^4T_4)\exp(z^5T_5)\nonumber\\
& & +\exp(z^1T_1)\exp(z^2T_2)\exp(z^3T_3)[dz^4T_4]\exp(z^4T_4)\exp(z^5T_5)\nonumber\\
& &+\exp(z^1T_1)\exp(z^2T_2)\exp(z^3T_3)\exp(z^4T_4)[dz^5T_5]\exp(z^5T_5)\Big). \label{gLeftform}
\end{eqnarray}
Since $T_1$ commutes with every generator, the first term of (\ref{gLeftform}) reduces to
\begin{equation}
dz^1T_1.
\end{equation}
The second term of  (\ref{gLeftform}) is
\begin{equation}
\exp(-z^5T_5)\exp(-z^4T_4)\exp(-z^3T_3)[dz^2T_2]\exp(z^3T_3)\exp(z^4T_4)\exp(z^5T_5).
\end{equation}
Since $T_4$ and $T_5$ commute with $T_2$ and $T_3$, the second term will be
\begin{equation}
\exp(-z^3T_3)[dz^2T_2]\exp(z^3T_3). \label{gsecond}
\end{equation}
The Baker-Campbell-Hausdorff formula is
\begin{eqnarray}
e^X Y e^{-X} = Y +[X,Y]+\frac{1}{2!}[X,[X,Y]]+\frac{1}{3!}[X,[X,[X,Y]]]+\cdots.
\end{eqnarray}
Using this, the term (\ref{gsecond}) can be written as
\begin{equation}
dz^2T_2-z^3dz^2[T_3,T_2]+\cdots. \label{gsecond2}
\end{equation}
Since the group is 2-step nilpotent Lie group, the terms with $[T_3,[T_3,T_2]]$ and higher will be zero. The result is
\begin{equation}
dz^2T_2+mz^3dz^2T_1.
\end{equation}
The third term of (\ref{gLeftform}) is
\begin{equation}
dz^3T_3
\end{equation}
while the fourth term is
\begin{equation}
\exp(-z^5T_5)[dz^4T_4]\exp(z^5T_5).
\end{equation}
By a similar reasoning  as that leading to (\ref{gsecond2}), this term becomes
\begin{equation}
dz^4T_4+mz^5dz^4T_1.
\end{equation}
The last term is 
\begin{equation}
dz^5T_5.
\end{equation}
Therefore, the left-invariant one-form is
\begin{equation}
g^{-1}dg = P^aT_a = (dz^1+mz^3dz^2+mz^5dz^4)T_1+(dz^2)T_2+(dz^3)T_3+(dz^4)T_4+(dz^5)T_5.
\end{equation}

\section{Appendix: Calculation of discrete  identifications of coordinates}
Let $\mathcal{G}_5$ be the five dimensional nilpotent Lie group with non-vanishing commutators
\begin{equation}
[T_2,T_3] = mT_1, \qquad [T_4,T_5] = mT_1.
\end{equation}
A group element $g$ can be written as
\begin{equation}
g = \exp(z^1T_1)\exp(z^2T_2)\exp(z^3T_3)\exp(z^4T_4)\exp(z^5T_5),
\end{equation}
where $z^1,\cdots, z^5$ are local coordinates on $\mathcal{G}_5.$
Let $\Gamma$ be the cocompact subgroup of $\mathcal{G}_5$ of group elements of the form
\begin{equation}
h = \exp(n^1T_1)\exp(n^2T_2)\exp(n^3T_3)\exp(n^4T_4)\exp(n^5T_5),
\end{equation}
where $n^i \in \mathbb{Z},$ and $ i = 1, \cdots, 5$.

Consider the left action of $h$ on $g$
\begin{eqnarray}
h \cdot g &=&  \exp(n^1T_1)\exp(n^2T_2)\exp(n^3T_3)\exp(n^4T_4)\exp(n^5T_5) \cdot\nonumber\\
& &\Big(\exp(z^1T_1)\exp(z^2T_2)\exp(z^3T_3)\exp(z^4T_4)\exp(z^5T_5)\Big).
\end{eqnarray}
Since $T_1$ commutes with every element  and $T_2$ and $T_3$ commute with $T_4$ and $T_5$, we get
\begin{eqnarray}
h\cdot g &=& \exp((z^1+n^1)T_1)\Big(\exp(n^2T_2)\exp(n^3T_3)\exp(z^2T_2)\exp(z^3T_3)\Big)\nonumber\\
& & \Big(\exp(n^4T_4)\exp(n^5T_5) \exp(z^4T_4)\exp(z^5T_5)\Big).
\end{eqnarray}
Consider the product
\begin{equation}
\exp(n^3T_3)\exp(z^2T_2). \label{product1}
\end{equation}
Using the product rule
\begin{equation}
e^X e^Y = e^{(Y+[X,Y]+\frac{1}{2!}[X,[X,Y]]+\frac{1}{3!}[X,[X,[X,Y]]]+\cdots )}e^X,
\end{equation}
the product (\ref{product1}) becomes
\begin{equation}
\exp(n^3T_3)\exp(z^2T_2) = \exp(z^2T_2-mn^3z^2T_1)\exp(n^3T_3).
\end{equation}
Then the product $h \cdot g$ becomes
\begin{eqnarray}
h\cdot g &=& \exp[(z^1+n^1-mn^3z^2-mn^5z^4)T_1]
\exp[(z^2+n^2)T_2]\exp[(z^3+n^3)T_3]\nonumber\\
& & \exp[(z^4+n^4)T_4]\exp[(z^5+n^5)T_5].
\end{eqnarray}
The quotient space $\mathcal{G}_5/\Gamma$ is obtained by identifying $g$ with  $h\cdot g$, 
\begin{equation}
g \sim h\cdot g.
\end{equation}
The  global structure of $\mathcal{G}_5/\Gamma$ is then specified by the following identification of local coordinates
\begin{eqnarray}
z^1 &\sim& z^1+n^1-mn^3z^2-mn^5z^4,\nonumber\\
z^2 &\sim& z^2+n^2,\nonumber\\
z^3 &\sim& z^3+n^3,\nonumber\\
z^4 &\sim& z^4+n^4,\nonumber\\
z^5 &\sim& z^5+n^5.
\end{eqnarray}


\section{Appendix: T-fold solutions}
The metric and $B$-field of the T-fold which is T-dual to the $T^2$ bundle over $T^4$ fibred over a line
is given by T-dualising the metric (\ref{T2T4metric}) in the $z^3$ direction. This results in:
\begin{equation}
g =     \left(\begin{array}{cccccc}A_5  & -C& 0 &  m  z^6 C& m  z^6 A_5& 0
\\
 -C& A_4& 0
  & 
  -m  z^6 A_4
  & -m  z^6 C
& 0
\\ 0 & 0 & \frac{1}{f}& 0 & 0 & 0
\\ m  z^6 C&
-m  z^6 A_4
& 0 & m^2\,{(z^6)}^2
-D_5
+1 & m^2 ( z^6)^2 C
& 0
\\
m  z^6 A_5
 &
  -m  z^6 C
   & 0 
   & m^2 ( z^6)^2 C
   &
    m^2\,{(z^6)}^2-
   D_4
    +1 & 0
    \\ 0 & 0 & 0 & 0 & 0 & 1 \end{array}\right),\label{T-fold1metric}
\end{equation}
\begin{equation}
B =     \left(\begin{array}{cccccc} 0 & 0 & \frac{m\,z^4}{f}& 0 & 0 & 0\\ 0 & 0 & \frac{m\,z^5}{f}& 0 & 0 & 0\\ -\frac{m\,z^4}{f}& -\frac{m\,z^5}{f}& 0 & \frac{m^2\,z^5\,z^6}{f}& -\frac{m^2\,z^4\,z^6}{f}& 0\\ 0 & 0 & -\frac{m^2\,z^5\,z^6}{f}& 0 & 0 & 0\\ 0 & 0 & \frac{m^2\,z^4\,z^6}{f}& 0 & 0 & 0\\ 0 & 0 & 0 & 0 & 0 & 0 \end{array}\right),\label{T-fold1Bfield}
\end{equation}
where 
$$f = {1+m^2\,{(z^4)}^2+m^2\,{(z^5)}^2} $$
and
\begin{equation}
A_5=\frac{m^2\,{(z^5)}^2+1}{f}, \qquad A_4=\frac{m^2\,{(z^4)}^2+1}{f}, \qquad C=\frac{m^2\,z^4\,z^5}{f}
\end{equation}
\begin{equation}
D_5=\frac{m^4\,{(z^5)}^2\,{(z^6)}^2}{f}
,\qquad
D_4= \frac{m^4\,{(z^4)}^2\,{(z^6)}^2}{f}
\end{equation}

The metric and $B$-field of the T-fold which is T-dual to the $T^3$ bundle over $T^4$ fibred over a line
is given by T-dualising the metric (\ref{T3T4metric}) in the $z^4$ direction. This results in:
\begin{equation}
g =     \left(\begin{array}{ccccccc} A_{5} & -C_{56} & -C_{57} & 0 & 0 & m\,z^7-D_{557} & 0\\ -C_{56} & A_{6} & -C_{67} & 0 & -m\,z^7 & -D_{567} & 0\\ 
-C_{57}& -C_{67} & A_{7} & 0 & m\,z^6 & -D_{577}& 0\\ 
0 & 0 & 0 & \frac{1}{f} & 0 & 0 & 0\\ 
0 & -m\,z^7 & m\,z^6 & 0 & m^2\,{(z^6)}^2+m^2\,{(z^7)}^2+1 & 0 & 0\\ 
m\,z^7-D_{557} & -D_{567} & -D_{577} & 0 & 0 & m^2\,{(z^7)}^2-\frac{m^4\,{(z^5)}^2\,{(z^7)}^2}{f}+1 & 0\\ 
0 & 0 & 0 & 0 & 0 & 0 & 1 \end{array}\right),\label{T-fold2metric}
\end{equation}
\begin{equation}
B =     \left(\begin{array}{ccccccc} 0 & 0 & 0 & \frac{m\,z^5}{f} & 0 & 0 & 0\\ 0 & 0 & 0 & \frac{m\,z^6}{f} & 0 & 0 & 0\\ 0 & 0 & 0 & \frac{m\,z^7}{f} & 0 & 0 & 0\\ -\frac{m\,z^5}{f} & -\frac{m\,z^6}{f} & -\frac{m\,z^7}{f} & 0 & 0 & -\frac{m^2\,z^5\,z^7}{f} & 0\\ 0 & 0 & 0 & 0 & 0 & 0 & 0\\ 0 & 0 & 0 & \frac{m^2\,z^5\,z^7}{f} & 0 & 0 & 0\\ 0 & 0 & 0 & 0 & 0 & 0 & 0 \end{array}\right),\label{T-fold2Bfield}
\end{equation}
where $$ f ={1 + m^2\,{(z^5)}^2+m^2\,{(z^6)}^2+m^2\,{(z^7)}^2}\,  , $$
$$A_5 = 1-\frac{m^2\,{(z^5)}^2}{f} , A_6 = 1-\frac{m^2\,{(z^6)}^2}{f} , A_7 = 1-\frac{m^2\,{(z^7)}^2}{f}\, , $$
$$C_{56} = \frac{m^2\,z^5\,z^6}{f} , C_{57} = \frac{m^2\,z^5\,z^7}{f}, C_{67} = \frac{m^2\,z^6\,z^7}{f} \, ,$$
$$D_{557} = \frac{m^3\,{(z^5)}^2\,z^7}{f}, D_{567} = \frac{m^3\,z^5\,z^6\,z^7}{f}, D_{577} = \frac{m^3\,(z^5)\,{(z^7)}^2}{f} \, . $$





\end{document}